\documentclass[letter,12pt]{article}
\usepackage{epsfig,amsmath,amssymb,verbatim,mathrsfs}
 \pdfoutput=1
\def\vector#1#2{\left( \begin{array}{c}#1\\ #2\end{array}\right)}
\def\beq{\begin{equation}}
\def\eeq{\end{equation}}
\def\bea{\begin{eqnarray}}
\def\eea{\end{eqnarray}}
\def\bmat{\begin{pmatrix}}
\def\emat{\end{pmatrix}}
\def\bei{\begin{itemize}}
\def\eei{\end{itemize}}
\def\eq#1{eq.\ (\ref{#1})}

\def\gev{\, {\rm GeV}}

\newcommand{\gsim}{\lower.7ex\hbox{$\;\stackrel{\textstyle>}{\sim}\;$}}
\newcommand{\lsim}{\lower.7ex\hbox{$\;\stackrel{\textstyle<}{\sim}\;$}}

\def\vuo{v_{u1}}
\def\vut{v_{u2}}
\def\vdo{v_{d1}}
\def\vdt{v_{d2}}

\def\vo{v_{u1}}
\def\vt{v_{d1}}
\def\vth{v_{u2}}
\def\vf{v_{d2}}

\def\bot{b_{12}}
\def\boo{b_{11}}
\def\bto{b_{21}}
\def\btt{b_{22}}
\def\btot{{b}_{12}}
\def\btoo{{b}_{11}}
\def\btto{{b}_{21}}
\def\bttt{{b}_{22}}
\def\muoo{\mu_{11}}
\def\muot{\mu_{12}}
\def\muto{\mu_{21}}
\def\mutt{\mu_{22}}

\def\mtdeluo{{b}_{11}\frac{v_{d1}}{v_{u1}}+b_{12}\frac{v_{d2}}{v_{u1}}}
\def\mtdelut{{b}_{21}\frac{v_{d1}}{v_{u2}}+b_{22}\frac{v_{d2}}{v_{u2}}}
\def\mtdeldo{{b}_{11}\frac{v_{u1}}{v_{d1}}+b_{21}\frac{v_{u2}}{v_{d1}}}
\def\mtdeldt{{b}_{12}\frac{v_{u1}}{v_{d2}}+b_{22}\frac{v_{u2}}{v_{d2}}}
\makeatletter
\renewcommand{\section}{\@startsection{section}{1}{0em}%
        {-3.25ex \@plus -1ex \@minus -.2ex}%
        {2.0ex \@plus.2ex}%
        {\normalfont\large\bfseries}}
\renewcommand{\subsection}{\@startsection{subsection}{2}{0em}%
        {-2.75ex\@plus -1ex \@minus -.2ex}%
        {1.25ex \@plus .2ex}%
        {\normalfont\bfseries}}
\renewcommand{\subsubsection}%
        {\@startsection{subsubsection}{3}{0em}%
        {-2.0ex\@plus -1ex \@minus -.2ex}%
        {1.0ex \@plus .2ex}%
        {\normalfont\itshape}}
\makeatother

\begin{document}
\baselineskip=18pt

\begin{titlepage}

\noindent
\begin{flushright}
MCTP-09-55 \\
CERN-PH-TH-2009-228\\
\end{flushright}
\vspace{1cm}

\begin{center}
  \begin{Large}
    \begin{bf}
    Next Generation Higgs Bosons: \\
    Theory, Constraints and Discovery \\
    Prospects at the Large Hadron Collider

        \end{bf}
  \end{Large}
\end{center}

\vspace{0.5cm}
\begin{center}
\begin{large}
Rick S. Gupta, James D. Wells \\
\end{large}

\vspace{0.3cm}
\begin{it}
CERN, Theoretical Physics, CH-1211 Geneva 23, Switzerland \\
\vspace{0.15cm}
Michigan Center for Theoretical Physics (MCTP) \\
        ~~University of Michigan, Ann Arbor, MI 48109-1120, USA \\
\vspace{0.1cm}
\end{it}

\vspace{1cm}
\end{center}

\begin{abstract}

Particle physics model building within the context of string theory suggests that further copies of the Higgs boson sector may be expected. Concerns regarding tree-level flavor changing neutral currents are easiest to allay if little or no couplings of next generation Higgs bosons are allowed to Standard Model fermions. We detail the resulting general Higgs potential and mass spectroscopy in both a Standard Model extension and a supersymmetric extension. We present the important experimental constraints from meson-meson mixing, loop-induced $b\to s\gamma$ decays and LEP2 direct production limits. We investigate the energy range of  valid perturbation theory of these ideas. In the supersymmetric context we present a class of examples that marginally aids the fine-tuning problem for parameter space where the lightest Higgs boson mass is greater than the Standard Model limit of 114 GeV.  Finally,  we study collider physics signatures generic to next generation Higgs bosons, with special emphasis on $Ah\to hhZ\to 4b+2l$ signal events, and describe the capability of discovery at the Large Hadron Collider.

\end{abstract}

\vspace{3cm}

\begin{flushleft}
\begin{small}
December 2009
\end{small}
\end{flushleft}

\end{titlepage}

%%%%%%%%%%%%%%%%%%%%%%%%%

\tableofcontents
\setcounter{page}{2}

%%%%%%%%%%%%%%%%%%%%%%%%%%%%%%%%%%
\vfill\eject

\section{Generations of Higgs Bosons}

Chiral matter comes in three generations. The simplest hypotheses of electroweak symmetry breaking and fermion mass generation assumes the existence of one Higgs boson in the case of the Standard Model and a pair of Higgs bosons in the case of supersymmetry. We ask here what the consequences are of having more generations of Higgs bosons in analogy to fermion matter content. We are not the first to ask this question and investigate answers (see for eg.~\cite{Ambroso:2008kb},~\cite{Z3 Munoz} and~\cite{sher} to be highlighted later). Some of our discussion will be known to readers, but that is only to set the stage for describing further material we have developed and in particular detailing Large Hadron Collider (LHC) implications for next generation Higgs boson ideas that survive scrutiny.

The question is of increased interest of late for two reasons. One, the LHC begins soon and enters the prime real estate of Higgs boson phenomenology, and we should be prepared to discover all reasonable and viable ideas. The physical particle spectrum of the Higgs sector, if it exists, is speculation at present. Investigating various scenarios that may yield phenomenology that is different from the simplest Standard Model (SM) approach is needed in order to develop more interpretive power over the data when it comes.  

A second reason to consider a next generation of Higgs bosons is from recent developments in string phenomenology.  Some approaches to particle physics model building from string theory suggest that further copies of Higgs bosons may be generic among solutions.  For example, in the work 
of~\cite{Ambroso:2008kb} a second Higgs generation is generic among the heterotic vacua, and may even be more copious than single generation Higgs boson theories. It is also typical in this approach that there is a selection rule that allows only the first generation of Higgs bosons to couple to the fermions. We will review later why this aspect is very helpful for the viability of a next generation of Higgs bosons.

Next generation Higgs bosons are motivated in other theories as well. For example, in theories with branes at singularities  bifundamental states come from the same quivers, and multiplicities of Higgs pairs are generic just like multiplicities of other representations. In intersecting $D$-brane theories, the chiral content is constrained by topological intersection numbers, but the vector-like states can be many-fold. Usually only our self-imposed restrictions in seeking solutions results in one generation. In heterotic orbifold models exotics are generic. Restrictions to three families of fermions rarely necessarily restricts Higgs bosons to one pair.  Some approaches, such as $Z_3$ orbifolds with two Wilson lines~\cite{Z3 Munoz}, naturally provide three generations of Higgs bosons, for example.

Most physicists nowadays carry the vague suspicions that additional Higgs bosons are disastrous unless introduced into very restricted frameworks. They give the photon a mass, or result in unacceptable tree-level flavor changing neutral currents. This is to a large degree correct, but there are interesting viable limiting cases, touched on above, that are supported by theory model building. We set out to elucidate some general conditions for the viability of next generation Higgs bosons. We detail a formalism for the analysis, including determining the mass matrices and mixing angles in both the SM and Supersymmetry. This culminates in a study of a key process at the LHC that is signal for next generation Higgs bosons.

%%%%%%%%%%%%%%%%%%%%%%%%%%%%%%%%%%%%%%%
\section{Overcoming Tree-Level Flavor Changing Neutral Currents}

Let us begin by considering an extra Higgs doublet $\Phi_{extra}$ that is added to the SM Higgs doublet $\Phi_{sm}$. The vacuum expectation values (vev) of each are $\langle \Phi\rangle=v_{sm}$ and $\langle \Phi_{extra}\rangle =v_{extra}$, subject to the condition that $v^2=v^2_{sm} +v_{extra}^2=(246\gev )^2$.  We assume that both Higgs doublets couple to the SM fermions. 
From these two doublets, three degrees of freedom are eaten and become longitudinal components of $W^\pm_L$ and $Z_L^0$, and five degrees of freedom are left: the scalar mass eigenstates $\{ H,h\}$, the pseudoscalar $A$, and the charged Higgs bosons $H^\pm$.

%%%%%%%%%%%%%%%%%%%%%%%%%%%%%%%%%%%%
% fcnc.tex

It is always possible to write the Yukawa Lagrangian terms as,
\bea
{\cal L}_Y &=&\frac{\sqrt{2}m^{U}_i}{v}\delta_{ij}\bar{Q}_{iL} \tilde{\Phi}_{vev} U_{jR}+\frac{\sqrt{2}m^{D}_i}{v}\delta_{ij}\bar{Q}_{iL} {\Phi}_{vev} D_{jR}+\frac{\sqrt{2}m^{E}_i}{v}\delta_{ij}\bar{L}_{iL} {\Phi}_{vev} E_{jR} \nonumber\\
& &+{\sqrt{2}}~\xi_{ij}^U \bar{Q}_{iL} \tilde{\Phi}_{\perp} U_{jR} +{\sqrt{2}}~\xi^D_{ij}\bar{Q}_{iL} {\Phi}_{\perp} D_{jR}+ {\sqrt{2}}~\xi^E_{ij}\bar{L}_{iL} {\Phi}_{\perp} E_{jR}+c.c.
\eea
where $E_i$, $U_i$ and $D_i$  are mass eigenstates of leptons, up type and down type quarks and $\xi^{U,D}_{ij}$ are \textit{a priori} arbitrary. The definition of $\Phi_{vev}$ is the linear combination that contains the full vev,
\beq
\Phi_{vev}=\frac{v_{extra}}{v}\Phi_{extra}+\frac{v_{sm}}{v}\Phi_{sm},~{\rm and}
\eeq
\beq
\Phi_{\perp}=\frac{v_{extra}}{v}\Phi_{sm}-\frac{v_{sm}}{v}\Phi_{extra}
\eeq
is the perpendicular state with no vev associated to it. In general, there is nothing to forbid the off-diagonal elements of $\xi_{ij}^{U,D,E}$ from being ${\cal O}(1)$. This is the origin of the tree-level Flavor Changing Neutral Curent (FCNC) problem of extra Higgs bosons.

To estimate the experimental upper bound on the off diagonal elements of $\xi^{U,D}_{ij}$ let us assume that the matrices $\xi^{U,D}$ are real and symmetric. We then obtain the following Feynman rules for the scalar and pseudoscalar mass eigenstates (the vertex factor being $-i$ times the expressions below),
\bea
H\bar U_iU_j, H\bar D_iD_j  & = & c_{\alpha'} \frac{m^{U,D}_i}{v}\delta_{ij}+s_{\alpha'} \xi_{ij}^{U,D} \\
h\bar U_iU_j, h\bar D_iD_j  & = & -s_{\alpha'} \frac{m^{U,D}_i}{v}\delta_{ij}+c_{\alpha'} \xi_{ij}^{U,D} \\
A\bar U_iU_j, A\bar D_iD_j & = & i \gamma_5 \xi_{ij}^{U,D}.
\label{FCcoupling}
\eea
Here the mixing angle $\alpha'$ is the one that rotates from  $\{ \Phi_{vev},\Phi_{\perp}\}$ to the mass eigenstates $\{ H,h\}$.  
\begin{figure}[t]
\hspace{-1.5in}
\includegraphics[width=1.5\columnwidth]{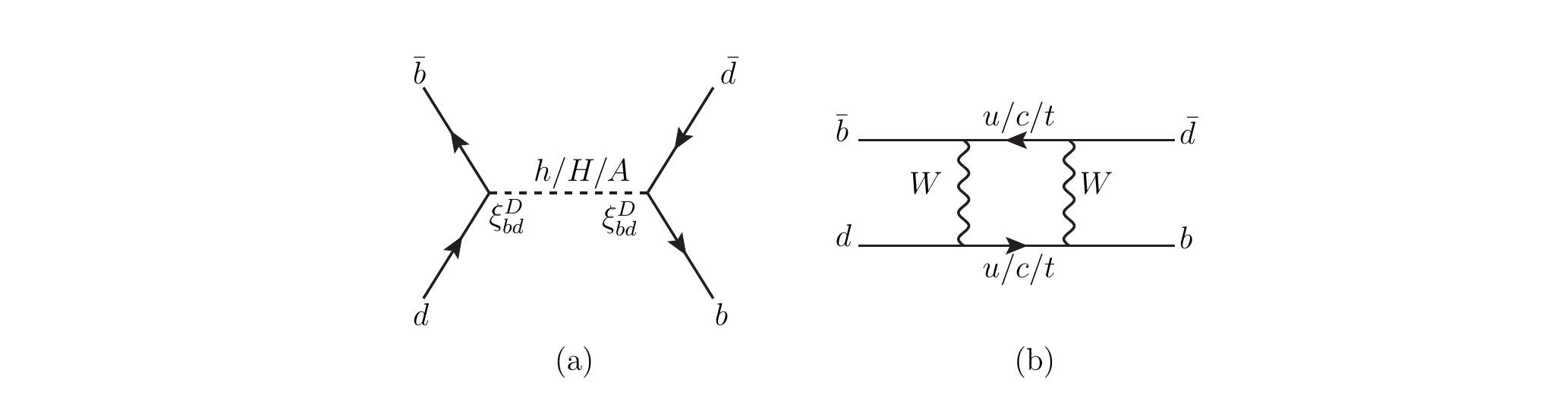}
\caption{Flavor changing neutral current contributions to $B_d^0-\bar B_d^0$ mixing from (a) Higgs exchange diagrams in an arbitrary 2HDM (there are also $t$-channel diagrams that we have not shown here), and (b) SM gauge contributions. Note that the SM diagrams are one-loop whereas the competing Higgs exchange is tree-level. Experiment is consistent with SM results, which implies severe constraints on the Higgs flavor-changing neutral current couplings $\xi_{ij}^F\ll 1$.}
\label{fig:feyndig}
\end{figure}
The most stringent constraints on $\xi_{ij}^{U,D}$ come from $F^0-\bar{F}^0$ mixing (where $F=K,B_d,D, B_s$). In a two-Higgs doublet model (2HDM)  with arbitrary Yukawa couplings there is a tree-level contribution to the $F^0-\bar{F}^0$ mass splitting because of  diagrams like Fig.~\ref{fig:feyndig}(a). For $\alpha^{'}=0$ using the expressions derived in~\cite{Atwood:1996vj} we get in the vacuum insertion approximation,
\beq
M_F \Delta M_F =  {\xi_{ij}^{U,D}}^2 \left(s_{\alpha'}^2\frac{S_F}{m_H^2}+c_{\alpha'}^2\frac{S_F}{m_h^2}-\frac{P_F}{m_A^2}\right)
\label{exprss}
\eeq
where,
\bea
S_F &=& \frac{B_F f_F^2 M_F^2}{6}\left(1+\frac{ M_F^2}{(m_i +m_j)^2}\right)\\
P_F &=& -\frac{B_F f_F^2 M_F^2}{6}\left(1+\frac{11 M_F^2}{(m_i +m_j)^2}\right).
\eea
Here $M_F$ is the mass of the meson, $m_H,m_h$ and $m_A$ are the masses of $H,h$ and $A$, $f_F$ is the pseudoscalar decay constant and $B_F$ is the $B$-parameter of the vacuum insertion approximation defined in \cite{Atwood:1996vj}. We present the values of these parameters and the  experimental values for $\Delta M_F$ in Table~\ref{table:mesons}.
\begin{table}[t]
\begin{tabular}{cccc}
\hline
Meson (quarks) & $B_F$ & $f_F$ (GeV) & $\Delta M_F^{\rm expt}$ (GeV) \\
\hline
$K^0\,( d\bar s)$  & $0.79\pm 0.04\pm0.08$ & 0.159&$(3.476\pm 0.006)\times 10^{-15} $~\cite{pdg}\\  
$D^0\,(\bar uc)$  & $0.82\pm0.01$ & 0.165 &$(0.95\pm0.37)\times 10^{-14}$ \\
$B_d^0\, (d \bar b)$  &$1.28\pm0.05\pm0.09$~\cite{Lubicz:2007yu} &$0.216\pm0.022$ &  $(3.337\pm0.033)\times 10^{-13} $~\cite{pdg}\\
$B_s^0\, ( s\bar b)$  &\small $-$&$0.281\pm0.021^*$  &$(117.0\pm0.8)\times 10^{-13} $~\cite{pdg}\\ 
\hline
\end{tabular}
\caption{Data associated with the neutral mesons $K^0$, $B^0_d$ and $D^0$. Values have been obtained from~\cite{Lunghi:2007ak} unless mentioned otherwise.
\newline
\newline \footnotesize*This is actually the value of the product $f_{B_s} \sqrt{B_{B_s}}$.}
\label{table:mesons}
\end{table}

The Cheng-Sher ansatz~\cite{Cheng:1987rs} is sometimes assumed for the flavor changing couplings,
\beq
\xi^{U,D,E}_{ij}=\lambda^{U,D,E}_{ij} \frac{\sqrt{m_i m_j}}{v},
\label{chengsher}
\eeq
as many approaches to flavor model building would give rise to it.
Let us now find the maximum allowed value of  $\xi^{U/D}_{ij}$  consistent with the experimental data and simultaneously determine the corresponding $\lambda^{U/D}_{ij}$. For  $B_{d}^0$ and $B_{s}^0$ we  use the SM predictions from lattice QCD, $\Delta M_{B_d}^{SM}=(4.5\pm1.0)\times 10^{-13}$ GeV ~\cite{Ball:2006xx} and $\Delta M_{B_s}^{SM}=(135\pm20)\times 10^{-13}$ GeV~\cite{Lunghi:2007ak}  and add the theoretical error in quadrature to the experimental error. To find the upper bound on $\xi^{D}_{db}$ and $\xi^{D}_{sb}$ we demand that the sum of the SM value and the 2HDM contribution from \eq{exprss}  does not exceed the experimetal value in Table~\ref{table:mesons} by more than two standard deviations. In obtaining the upper bounds on $\xi^{D}_{ds}$ and $\xi^{U}_{uc}$ we require that just the 2HDM contribution from \eq{exprss} does not exceed the experimetal value in Table~\ref{table:mesons} by more than two standard deviations. Taking  $m_H=m_h=m_A=120$ GeV  we  find,
\bea
\xi^{D}_{ds}&\lesssim & 1\times 10^{-5}~~~~~~~~~~\lambda^{D}_{ds}\lesssim 0.1\\
\xi^{D}_{uc}&\lesssim & 3\times 10^{-5}~~~~~~~~~~\lambda^{D}_{uc}\lesssim 0.2\\
\xi^{D}_{db}&\lesssim & 4\times 10^{-5}~~~~~~~~~~\lambda^{D}_{db}\lesssim 0.06\\
\xi^{D}_{sb}&\lesssim & 2\times 10^{-4}~~~~~~~~~~\lambda^{D}_{sb}\lesssim 0.06.
\eea 
The upper bounds stated above have an uncertainty due to the uncertainty in the value of $f_{F} \sqrt{B_{F}}$. From \eq{exprss} we can see that a 10$\%$ uncertainty in $f_{F} \sqrt{B_{F}}$ would translate to a 10$\%$ uncertainty in the upper bound.  Greater precision in lattice estimates of the SM values is required for more stringent constraints on $\xi^{U/D}_{ij}$ from $F^0-\bar{F}^0$ mixing. 

Future measurement of branching ratios  of rare $B$ decay modes  such as $B(\bar{B}_s \rightarrow \mu^+ \mu^-)$  at LHCb is another way  effects of the flavor changing couplings $\xi^{D}_{bs}$ and $\xi^{D}_{sb}$ may be discovered. Expressions for this branching ratio in the SM, and the Higgs contribution in a 2HDM with arbitray Yukawa couplings, can be found in Ref.~\cite{Joshipura:2009ej}. The SM value is $(3.51\pm 0.50)\times 10^{-9}$~\cite{Joshipura:2009ej} and the current Tevatron upper bound at 95$\%$ CL is $5.8\times10^{-8}$~\cite{:2007kv}. As LHCb would reach the sensitivity to measure the SM value~\cite{Bettler:2009rf}, we can estimate the kind of upper bounds LHCb experiments would put on $\xi^{D}_{bs}$ and $\xi^{D}_{sb}$ by demanding that the Higgs contribution to this braching ratio is less than the SM value. Assuming again that  $\xi^{D}_{bs}$ and $\xi^{D}_{sb}$ are real and equal to each other, and that $\xi_{ij}=\lambda_{ij}\sqrt{m_im_j}/v$, we obtain
\bea
\lambda^{D}_{sb} &\lesssim & 2\, \frac{(m_A/120~{\rm GeV})^2}{\lambda^{E}_{\mu \mu}}~~~~~~~{\rm (Tevatron~inferred~bound)}\\
\lambda^{D}_{sb} &\lesssim  & 0.6\, \frac{(m_A/120~{\rm GeV})^2}{\lambda^{E}_{\mu \mu}}~~~{\rm (LHCb~expected~sensitivity)}.
\eea
Although this limit from Tevatron and future sensitivity expectations at LHCb do not appear to be as powerful as meson-meson mixing constraints, the uncertainty in what $\lambda_{\mu\mu}^E$ might be in the presence of next generation Higgs bosons suggests that it should stay under consideration.
%If we take $\lambda^{E}_{\mu \mu}=\tan \beta$, its MSSM value,  we get,
%\bea
%\xi^{D}_{bs} &\lesssim& 7.10^{-4}\frac{(m_A/120~{\rm GeV})^2}{(\tan \beta/10)}~~~~~~~~{\rm (Tevatron)}\\
%\xi^{D}_{bs} &\lesssim& 2.10^{-4}\frac{(m_A/120~{\rm GeV})^2}{(\tan \beta/10)}~~~~{\rm (LHCb~expected)}.
%\eea
For a more thorough discussion of $B_s^0-\bar{B}_s^0$ mixing constraints and the processes $\bar{B}_s \rightarrow \mu^+ \mu^-,\bar{B}_d \rightarrow \bar{K}\mu^+ \mu^-$ see Ref.$~\cite{Joshipura:2009ej}$. For a discussion on the constraints on the diagonal couplings $\xi^{F}_{ii}$ see Ref.~\cite{Mahmoudi:2009zx}.

%%%%%%%%%%%%%%%%%%%%%%%%%%%%%%%%%%%%
We have seen above that the off-diagonal couplings $\xi^F_{ij}$ are required to be extremely small in order to satisfy FCNC constraints. There is a general class of solutions to this problem~\cite{Wells:2009kq} while admitting the existence of extra Higgs bosons in the spectrum. Tree-level FCNCs do not arise if Higgs boson interactions with the fermions take the form
\begin{eqnarray}
\Delta {\cal L}_{f}=y^D_{ij}\bar Q'_{iL}\,  F_d(\{ \Phi_k\})\, D'_{jR} + y^U_{ij}\bar Q'_{iL}\, F_u(\{ \Phi_k\}) \, U'_{jR} +
y^E_{ij}\bar L'_{iL}\, F_e(\{ \Phi_k\})\, E'_{jR} + c.c.
\label{no fcnc} 
\end{eqnarray}
where all the quark fields are gauge eigenstates. $F_{u,d,e}(\{\Phi_k\})$ are functions of Higgs fields $\{ \Phi_k\}$, constrained only by the requirements that they are independent of the fermionic flavor indices $i,j$ and that $F_u$ transforms like an $SU(2)_L$ doublet with hypercharge $-1/2$, and $F_d$ and $F_e$ transform like $SU(2)_L$ doublets with hypercharge $1/2$.

The generalized form of \eq{no fcnc} subsumes many ideas already present in the literature. For example, the SM Higgs sector is $F_u=H^c_{SM}$ and $F_d=F_e=H_{SM}$. The type II~\cite{Higgs type models} 2HDM~\cite{HHG} is $F_u=H_u$ and $F_d=F_e=H_d$. The type I 2HDM~\cite{HHG} is $F_u=\Phi^c_1$ and $F_d=F_e=\Phi_1$ with an additional $\Phi_2$ that does not couple to fermions. The leptophilic Higgs model of Ref.~\cite{Su:2009fz} is $F_u=\phi^c_q$, $F_d=\phi_q$ and $F_e=\phi_l$. 

There are an infinite variety of models that can satisfy \eq{no fcnc}. However, principles are expected to be at work to fall into this class if there is more than one Higgs boson. In the case of supersymmetry, the type II structure follows from holomorphy of the superpotential. In the case of type I theories, it is usually assumed that the second Higgs has, for example,  a discrete $Z_2$ symmetry $\Phi_2=-\Phi_2$ that forbids its direct coupling to fermions whereas $\Phi_1$ does not. 

The summary point is that a next generation Higgs boson is unlikely to satisfy FCNC currents due to tree-level mediated interactions unless a principle is invoked the ensures that it will satisfy the condition of \eq{no fcnc}. The most straightforward principle that we can invoke, and one that has been nicely illustrated recently in the model of~\cite{Ambroso:2008kb}, is a selection rule that forbids the next generation Higgs boson from coupling to fermions.  

%%%%%%%%%%%%%%%%%%%%%%%%%%%%%%%%%%%%%%%%%
\section{Next Generation Higgs bosons of Supersymmetry}

We begin with a discussion of next generation Higgs bosons in supersymmetry. In minimal supersymmetry there are already two Higgs doublets present in the spectrum. In unrestricted field theory, two Higgs doublets with arbitrary couplings are a disaster for flavor changing neutral currents. However, as mentioned above, supersymmetric theories have the special property that all superpotential operators must be holomorphic in the superfields.  Thus, it is impossible to employ one Higgs field to give masses to both up-type fermions and down-type fermions. The introduction of the second Higgs doublet solves this problem, and holomorphy is the principle by which tree-level FCNC's are held under control. The interaction lagrangian takes the form of the type II 2HDM in the form of \eq{no fcnc}.  Of course, when supersymmetry is broken, non-holomorphic interactions can induce additional couplings leading ultimately to potentially interesting flavor changing neutral currents~\cite{Babu:1999hn}, but those are naturally small over much of parameter space.

If we wish to add more Higgs doublets to supersymmetry, we must do it in up- and down-Higgs boson pairs. This is in sympathy with adding a new generation -- a new copy -- of the $\{H_u,H_d\}$ pair. More importantly, it is required to straightforwardly satisfy anomaly constraints due to the presence of new fermions that are the superpartners of the Higgs bosons.  When a next generation of Higgs doublets is added to the spectrum, holomorphy is no longer powerful enough to save us from tree-level FCNC.  Additional Yukawa couplings generally create FCNC challenges~\cite{Z3 Munoz,Mahmoudi:2009zx}.  In Ref.~\cite{sher}  supersymmetric three generation Higgs models have been considered where an exact symmetry prevents the next generation Higgs Bosons from coupling to the fermions. They show that these next generation Higgs bosons would not couple to the  standard Higgs bosons and under the assumption of equal scalar masses at unification scale they would not get vevs, thus avoiding FCNC problems. In general the simplest way out of the FCNC challenge is to assume that the extra Higgs doublet pairs couple very weakly, or not at all, to the fermions. This is the assumption we shall adopt for now.

The result of the discussion above is that we are overlaying a type I Higgs structure to our type II supersymmetric theory.  In other words, we are adding a Higgs boson pair that does not couple to the fermions on top of a Higgs pair that does. Our emphasis in this study is on the type I aspect (i.e., the next generation), and as such the supersymmetric type II feature is of less immediacy. Thus, we shall postpone a detailed discussion of additional constraints to later, to the less complex model of adding one Higgs boson to the SM. Nevertheless, we wish to present the mass eigenstates and mixing, and in particular highlight what effect multiple Higgs bosons have on the mass of the lightest Higgs boson scalar of supersymmetry.

\subsection{General Higgs Potential}

For the supersymmetric two Higgs generation model (SUSY 2HGM), we consider two additional Higgs superfields $H_{u2}$ and $H_{d2}$ charged under $SU(2)_L\times U(1)_Y$ just as $H_{u1}, H_{d1}$  respectively. The terms in the superpotential involving these superfields are,~\label{susy}
\beq
W=\mu_{11}H_{u1}H_{d1}+\mu_{12}H_{u1}H_{d2}+\mu_{21}H_{u2}H_{d1}+\mu_{22}H_{u2}H_{d2}.
\label{eq:super}
\eeq
From now on by $H_{u1}, H_{d1}, H_{u2}, H_{d2}$ we will denote only the scalar part of the superfields which have the components,
\begin{equation}
H_{ui}=\left( \begin{array}{c} H_{ui}^+ \\H^0_{ui}\end{array} \right)~~~~~~~~H_{di}=\left( \begin{array}{c} H^0_{di} \\H_{di}^-\end{array} \right).
\end{equation}
%\section{D-terms}
The $D$-terms are given by,
\begin{equation}
D^a=-g\sum_i(H_{ui}^\dagger T^a H_{ui}+H_{di}^\dagger T^a H_{di}).
\end{equation}
The $D$-term contribution to the potential is given by,
\bea
V_D &=& \frac{1}{2} \sum_a D_a D_a
\nonumber\\
&=&\lambda  [\sum_i(|H_{ui}|^2-|H_{di}|^2)]^2
  +\frac{g^2}{2}(\sum_i |(H_{ui}^{+*}H^0_{ui}+H^{0*}_{di} H_{di}^{-})|^2
  \nonumber\\
  & & 
 -\sum_i (|H^0_{ui}|^2-|H^0_{di}|^2)\sum_j(|H^+_{uj}|^2-|H^-_{dj}|^2)).
\label{eq:Dterm}
\eea
where $\lambda=\frac{g^2+g^{\prime 2}}{8}$.
%\section{F-Terms}
Using the form of the superpotential in \eq{eq:super} we get the $F$-term contribution,
\bea
V_F &=& (|\mu_{11}|^2+|\mu_{12}|^2 )|H_{u1}|^2 + (|\mu_{22}|^2+|\mu_{21}|^2 )|H_{u2}|^2 + (|\mu_{11}|^2+|\mu_{21}|^2 )|H_{d1}|^2\nonumber\\
& &  +(|\mu_{22}|^2+|\mu_{12}|^2 )|H_{d2}|^2 + (a_u H_{u1}^\dagger H_{u2} + a_d H_{d1}^\dagger H_{d2} + c.c.)
\label{fterm}
\eea
where,
\bea
a_d & = & \muoo^*\muot+\muto^*\mutt \\
a_u & = & \muoo^*\muto+\muot^*\mutt.
\eea
%\section{Soft SUSY breaking terms}

In addition to the $D$-terms and $F$-terms there are the soft SUSY breaking terms,
\bea
V_{soft} &=& m_{u1}^{'2} |H_{u1}|^2+m_{u2}^{'2} |H_{u2}|^2+m_{d1}^{'2} |H_{d1}|^2+m_{d2}^{'2} |H_{d2}|^2 +(\boo H_{u1} H_{d1}+ \nonumber\\
& & \bot H_{u1} H_{d2} +\bto H_{u2} H_{d1}+\btt H_{u2} H_{d2} + c.c.)
\label{soft}
\eea
where $H_{ui} H_{dj}=H^+_{ui} H^-_{dj}-H^0_{ui} H^0_{dj}$.
%\section{Electroweak symmetry breaking and mass matrices}

Adding all the terms from \eq{eq:Dterm},~(\ref{fterm}) and~(\ref{soft}) we get finally,
\begin{eqnarray}
V &=& V_D + V_F + V_{soft}
\nonumber\\
  &=& m_{u1}^2|H_{u1}|^2+m_{u2}^2|H_{u2}|^2+m_{d1}^2 |H_{d1}|^2+m_{d2}^2|H_{d2}|^2 
  + (a_u H_{u1}^\dagger H_{u2} + a_d H_{d1}^\dagger H_{d2} + c.c.) \nonumber\\
  & &+(\boo H_{u1} H_{d1}+ 
\bot H_{u1} H_{d2} +\bto H_{u2} H_{d1}+\btt H_{u2} H_{d2} + c.c.)+\lambda  [\sum_i(|H_{ui}|^2-|H_{di}|^2)]^2
 \nonumber\\
  & &  +\frac{g^2}{2}(\sum_i |(H_{ui}^{+*}H^0_{ui}+H^{0*}_{di} H_{di}^{-})|^2
 -\sum_i (|H^0_{ui}|^2-|H^0_{di}|^2)\sum_j(|H^+_{uj}|^2-|H^-_{dj}|^2)).
 \label{susypot}
\end{eqnarray}
Here  $m_i^2$ is the sum of $m_i^{'2}$ and the $F$-term contribution from \eq{fterm}. All the couplings  are required to be real by hermiticity with the exception of the bilinear couplings $b_{ij}$ and $a_i$ which are in general complex. By redefining the phases of the doublets $H_{u1}, H_{d2}$ and $H_{u2}$ only three of these six couplings can be chosen to be real so that the theory is in general CP violating.

Note that we are allowing $H_{u1,d1}$ to mix arbitrarily with $H_{u2,d2}$ in the Higgs potential only. Some mixing between the two generations is necessary as we do not want to have a global symmetry under which   the next generation Higgs bosons can be rotated with respect to the first generation because this would  lead to the presence of a massless Goldstone boson. When it comes to interacting with the fermions, as discussed before and emphasized again later,  next generation Higgs bosons are generally barred from having couplings unlike the first generation Higgs bosons.

\subsection{Electroweak symmetry breaking and scalar mass matrices}

We put $\langle H^+_{ui}\rangle =\langle H^-_{di}\rangle= 0$, $\langle H^0_i\rangle =\frac{1}{\sqrt{2}}v_i$ and demand that the first derivatives of  the potential with respect to the fields $H^0_i$ vanish to obtain,
\begin{equation}
 m_{d1}^2 -\tilde{b}_{11} \frac{v_{u1}}{v_{d1}}-\tilde{b}_{21} \frac{v_{u2}}{v_{d1}}+\tilde{a}_d \frac{v_{d2}}{v_{d1}}+\lambda (\sum_j v_{dj}^2-\sum_j v_{uj}^2)=0
 \label{min1}
\end{equation}
\begin{equation}
 m_{u1}^2 -\tilde{b}_{11} \frac{v_{d1}}{v_{u1}}-\tilde{b}_{12} \frac{v_{d2}}{v_{u1}}+\tilde{a}_u \frac{v_{u2}}{v_{u1}}-\lambda (\sum_j v_{dj}^2-\sum_j v_{uj}^2)=0
 \label{min2}
\end{equation}
\begin{equation}
 m_{d2}^2 -\tilde{b}_{22} \frac{v_{u2}}{v_{d2}}-\tilde{b}_{12} \frac{v_{u1}}{v_{d2}}+\tilde{a}_d \frac{v_{d1}}{v_{d2}}+\lambda (\sum_j v_{dj}^2-\sum_j v_{uj}^2)=0
 \label{min3}
\end{equation}
\begin{equation}
 m_{u2}^2 -\tilde{b}_{22} \frac{v_{d2}}{v_{u2}}-\tilde{b}_{21} \frac{v_{d1}}{v_{u2}}+\tilde{a}_u \frac{v_{u1}}{v_{u2}}-\lambda (\sum_j v_{dj}^2-\sum_j v_{uj}^2)=0.
 \label{min4}
\end{equation}
Here $\tilde{b}_{ij}={\rm Re~}b_{ij}$ and $\tilde{a}_i={\rm Re~}a_{i}$. It is necessary that $\langle H^+_{ui}\rangle =\langle H^-_{di}\rangle= 0$ for electromagnetism to remain unbroken. We can always choose one of the charged fields, say $H^-_{d1}$ to have an expectation value $\langle H^-_{d1}\rangle=0$ using the $SU(2)_L$ gauge freedom. To ensure that  $\langle H^+_{u1}\rangle=\langle H^-_{d2}\rangle=\langle H^+_{u2}\rangle=0$ is consistent with the minimization conditions, however, we must demand that in addition to eqs.~(\ref{min1})-(\ref{min4}) the second derivatives of the potential with respect to the charged fields at this point are positive. This is equivalent to demanding that the masses of the three physical charged Higgs bosons are positive. 

Also note that we have assumed  the vev of the neutral components $v_{ui}, v_{di}$ to be real and positive. One of the vevs say, $v_{d1}$, can be chosen to be real and positive using the $U(1)_Y$ gauge freedom. As we discussed earlier this theory is in general CP violating therefore we can choose the other vevs $\vuo$, $\vut$ and $\vdo$ to be real and positive simply by a convenient choice of phases of the respective doublet fields. These phases can then be absorbed in the bilinear couplings ${b}_{ij}$ and $a_i$ and  a redefinition of the quark fields. If the underlying theory ensures that all the bilinear couplings ${b}_{ij}$ and $a_i$ are real so that the Lagrangian conserves CP, in order to avoid spontaneous CP violation $\vuo$, $\vut$ and $\vdo$ must be real too. Therefore the point  $\langle A_{u1}\rangle=\langle A_{d2}\rangle=\langle A_{u2}\rangle=0$  must be a minima where $A_i=\sqrt{2}~ {\rm Im~}(H^0_i)$. For this,  in addition to eqs.~(\ref{min1})-(\ref{min4}) the second derivatives  of the potential with respect to the pseudoscalar fields $A_i$  must be positive also at this point. This is equivalent to demanding that the three physical pseudoscalars in the theory must have positive masses. Once $\vuo$, $\vut$ and $\vdo$ are known to be real they can always be chosen to be positive by a convenient choice of signs of the respective doublet fields, which can then be absorbed in  the the bilinear couplings ${b}_{ij}$ and $a_i$ and  a redefinition of the quark fields. If ${b}_{ij}$ and $a_i$ are complex the CP even states mix with the pseudoscalar states. To avoid this unnecessary complication to our present purposes we will present below the mass matrices assuming that the ${b}_{ij}$ and $a_i$  are real.

Using eqs.~(\ref{min1})-(\ref{min4}) we can eliminate the $m_i^2$. We can find the mass matrix of the four  CP-even scalars and the four pseudoscalars by substituting  $H^0_{i}=\frac{1}{\sqrt {2}}(v_{i}+h_{i}+ i A_{i})$. The mass matrix for the CP-even scalars, ${\cal M}^2_{H}$, in the  $\left\{d_1,u_1,d_2,u_2\right\}$ basis is,
\begin{displaymath}
\left(
  \begin{array}{cccc}
m^2_{H11} & -\btoo-2\lambda \vuo \vdo & {a}_d+2\lambda\vdo\vdt & -\btto-2\lambda\vut\vdo \\
-\btoo-2\lambda \vuo\vdo & m^2_{H22} & -\btot-2\lambda \vuo\vdt  & {a}_u+2\lambda\vuo\vut \\ 
{a}_d+2\lambda\vdo\vdt & -b_{12}-2\lambda\vuo\vdt  & m^2_{H33} & -b_{22}-2\lambda \vut\vdt \\
 -\btto-2\lambda\vut\vdo  & {a}_u+2\lambda\vuo\vut & -b_{22}-2\lambda\vut\vdt & m^2_{H44}
\end{array}\right)
\end{displaymath}
where,
\begin{eqnarray*}
m^2_{H11}&=&\mtdeldo-{a}_d \frac {v_{d2}}{v_{d1}}+2\lambda \vdo^2\\
m^2_{H22}&=&\mtdeluo-{a}_u \frac {v_{u2}}{v_{u1}}+2\lambda\vuo^2\\
m^2_{H33}&=&\mtdeldt-{a}_d \frac {v_{d1}}{v_{d2}}+2\lambda\vdt^2 \\
m^2_{H44}&=&\mtdelut-{a}_u \frac {v_{u1}}{v_{u2}}+2\lambda\vut^2.
\end{eqnarray*}
\noindent
The pseudoscalar mass matrix, ${\cal M}^2_{A}$, in the $\left\{d_1,u_1,d_2,u_2\right\}$ basis is,
\begin{displaymath}
\left( \begin{array}{cccc}
m^2_{A11}& \btoo & {a}_d & \btto\\
\btoo & m^2_{A22} & \btot & {a}_u \\ 
{a}_d & b_{12}  & m^2_{A33} & b_{22}\\
 \btto & {a}_u & b_{22}&m^2_{A44}
\end{array}\right) \nonumber
\end{displaymath}
where,
\begin{eqnarray*}
m^2_{A11}&=&\mtdeldo-{a}_d \frac {v_{d2}}{v_{d1}} \\
m^2_{A22}&=&\mtdeluo-{a}_u \frac {v_{u2}}{v_{u1}}\\
m^2_{A33}&=&\mtdeldt-{a}_d \frac {v_{d1}}{v_{d2}}\\
m^2_{A44}&=&\mtdelut-{a}_u \frac {v_{u1}}{v_{u2}}.
\end{eqnarray*}
The charged Higgs mass matrix differs from the pseudoscalar mass matrix only due to the last term in the potential in \eq{susypot}. In the basis  $\left\{H^{-*}_{d1},H^+_{u1},H^{-*}_{d2},H^+_{u2}\right\}$,
\begin{displaymath}
{\cal M}_{+}^2= {\cal M}_{A}^2+\frac{g^2}{4}\left( \begin{array}{cccc}
v_{u1}^2+v_{u2}^2-v_{d2}^2& v_{d1}v_{u1} &  v_{d1}v_{d2} &  v_{d1}v_{u2}\\
 v_{u1}v_{d1} & v_{d1}^2+v_{d2}^2-v_{u2}^2 & v_{u1}v_{d2}&  v_{u1}v_{u2} \\ 
 v_{d2}v_{d1} & v_{d2}v_{u1} & v_{u1}^2+v_{u2}^2-v_{d1}^2 &  v_{d2}v_{u2}\\
  v_{u2}v_{d1} &  v_{u2}v_{u1} &  v_{u2}v_{d2}& v_{d1}^2+v_{d2}^2-v_{u1}^2
\end{array}\right). 
\end{displaymath}
\subsection{Upper bound on the mass of the lightest CP even Higgs}

We want to transform the mass matrices above to the Runge basis which is defined as follows. One of the basis vectors in the Runge basis is,
\beq
\vec{V}_1=v_{d1}/v ~H_{d1}^c + v_{u1}/v~ H_{u1} + v_{d2}/v ~H_{d2}^c+ v_{u2}/v ~H_{u2}.
\label{vec1}
\eeq
Here $v=\sqrt{v_{u1}^2+v_{d1}^2+v_{u2}^2+v_{d2}^2}$. $H_{di}^c$ is in the $SU(2)_L$ conjugate representation and is given by,
\begin{equation}
H^c_{di}=\left( \begin{array}{c}-H_{di}^-  \\ H_{di}^{0*}\end{array} \right).
\label{cc}
\end{equation}
We choose the other basis vectors so that they are orthogonal to this vector and to each other. The simplest choices for two of the basis vectors are,
\bea
\vec{V}_2 &=& v_{u1}/v_1 ~H_{d1}^c - v_{d1}/v_1 ~H_{u1}\nonumber\\
\vec{V}_3 &=& v_{u2}/v_2 ~H_{d2}^c - v_{d2}/v_2 ~H_{u2}
\label{vec23}
\eea
where $v_1=\sqrt{v_{u1}^2+v_{d1}^2}$ and $v_2=\sqrt{v_{u2}^2+v_{d2}^2}$. We can find the fourth basis vector which is orthogonal to the first three by the expression,
\begin{eqnarray}
\vec{V'}_4 &=&  (\vec{V}_1.\vec{U})\vec{V}_1 + (\vec{V}_2.\vec{U})\vec{V}_2 + (\vec{V}_3.\vec{U})\vec{V}_3-\vec{U}
\nonumber\\
\vec{V}_4  &=& \frac {\vec{V'}_4}{\sqrt{\vec{V'}_4.\vec{V'}_4}}
\label{vec4}
\end{eqnarray}
where $\vec{U}$ can be any arbitrary vector. The transformation matrix is an $SO(4)$ rotation matrix,
\beq
R=( \vec{V_1}~ \vec{V_2} ~\vec{V_3} ~\vec{V_4}).
\label{rot}
\eeq

\begin{figure}[t]
%\hspace{1 cm}
\begin{center}
\includegraphics[width=0.75\columnwidth]{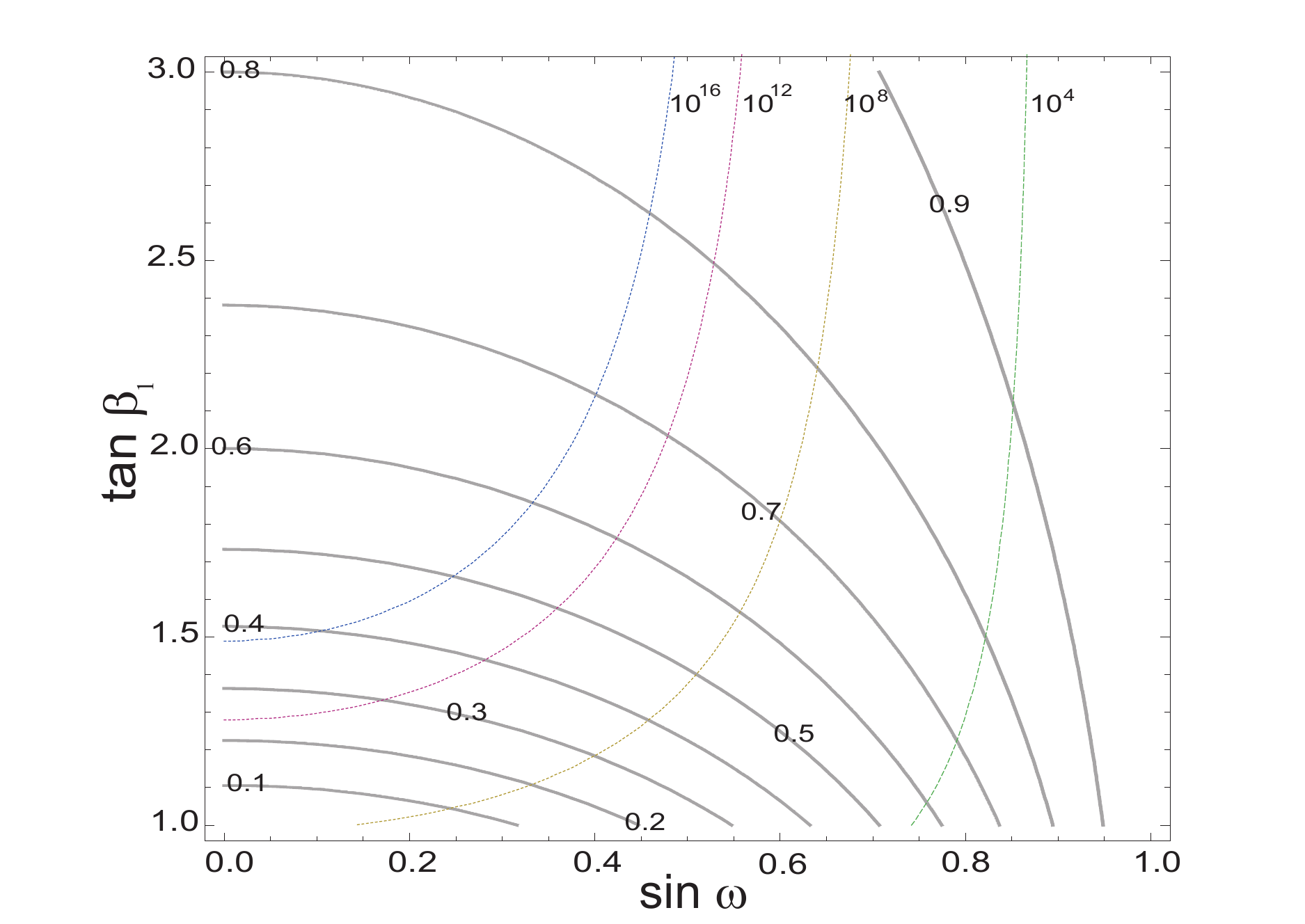}
\end{center}
\caption{Contours of the upper bound on the mass of the lightest Higgs (in $M_Z$ units)  in the SUSY 2HGM  taking $\cos 2 \beta_2=-1$. We have also shown contours of equal $\Lambda_{SC}$ (in GeV), the energy scale at which the top  Yukawa coupling $\lambda_t$ becomes larger than $4\pi$. Details about the calculation of $\Lambda_{SC}$  appear in section~\ref{secRG}.}
\label{fig:contour}
\end{figure}

%%%%%%%%%%%%%%%%%%%%%%%%%%

In the Runge basis, the mass matrices of the pseudoscalar, charged Higgs and CP-even scalar Higgs boson masses take on special form. For the pseudoscalar and charged Higgs mass matrix, the first row and first column contain all zeros, which is expected since the first basis vector $V_1$ is the  ``electroweak vev multiplet" which has all the vev. Thus, the CP-odd and charged components of the first basis vector in the Runge basis are the spin-zero Goldstone boson states absorbed by the $Z^0_L, W^+_L,$ and $W_L^-$ vector bosons. What remains is one more component of a full doublet, namely the CP-even part. We shall write the full CP-even mass matrix below and identify the matrix element corresponding to the mass of this CP even scalar and comment on its meaning.

%%%%%%%%%%%%%%%%%%%%%%%%%%

The CP even mass matrix in the Runge basis is,
\beq
{\cal M}^{'2}_{H} = R^{T}{\cal M}^2_{H} R.
\eeq
We get ${\cal M}^{'2}_{H}=$
\begin{small}
\begin{displaymath}
\left(\begin{array}{cccc}
 \frac{2\lambda(\sum_i (v_{di}^2-v_{ui}^2))^2}{v^2}&\frac{4 \lambda v_{d1} v_{u1}\sum_i (v_{di}^2-v_{ui}^2)}{v_{1} v} & \frac{4 \lambda v_{d2} v_{u2}\sum_i (v_{di}^2-v_{ui}^2)}{v_{2} v} &  \frac{4\lambda(v^2_{u2} v^2_{d1}-v^2_{u1} v^2_{d2})\sum_i (v_{di}^2-v_{ui}^2)}{v_1 v_2 v^2}\\
  \frac{4 \lambda v_{d1} v_{u1}\sum_i (v_{di}^2-v_{ui}^2)}{v_{1} v} & m^{'2}_{H22} & m^{'2}_{H23}& m^{'2}_{H24} \\
  \frac{4 \lambda v_{d2} v_{u2}\sum_i (v_{di}^2-v_{ui}^2)}{v_{2} v}  & m^{'2}_{H23}  & m^{'2}_{H33} & m^{'2}_{H34}\\
 \frac{4\lambda(v^2_{u2} v^2_{d1}-v^2_{u1} v^2_{d2})\sum_i (v_{di}^2-v_{ui}^2)}{v_1 v_2 v^2}& m^{'2}_{H24} & m^{'2}_{H34}& m^{'2}_{H44}
\end{array}\right) \nonumber
\end{displaymath}
\end{small}
where,
\begin{small}
\begin{eqnarray*}
m^{'2}_{H22} &=&  \frac{8 \lambda (v_{u1} v_{d1})^3 +\btoo v_1^4 +\btto \vuo^3 \vut-{a}_u \vdo^3 \vuo-{a}_d \vuo^3 \vdt+ \btot \vdo^3 \vdt}{v_1^2 \vuo \vdo}
\nonumber\\
m^{'2}_{H23} &=& \frac{{a}_d \vo \vth+\btot \vt \vth +\btto \vo \vf + {a}_u \vt \vf + 8 \lambda \vo \vth \vf}{v_1 v_2} 
\nonumber\\
m^{'2}_{H24} &=& \frac{v^2 (\btto \vo \vth+ {a}_u \vt \vth-{a}_d \vo \vf - \btot \vt \vf) + 8 \lambda (\vo \vt^3 \vth^2-\vo^3 \vt \vf^2)}{v v_1^2 v_2}
\nonumber\\
m^{'2}_{H33} &=&  \frac{8 \lambda (\vth \vf)^3 +\bttt v_2^4 +\btto \vt \vf^3 -{a}_u \vo \vf^3-{a}_d \vt \vth^3+ \btot \vo \vth^3}{v_2^2 \vut \vdt} 
\nonumber\\
m^{'2}_{H34} &=& \frac{v^2 ( \btto \vt \vf- {a}_u \vo \vf+{a}_d \vt \vth-\btot \vo \vth) + 8 \lambda( \vt^2 \vth^3 \vf -\vo^2\vth \vf^3 )}{v v_1 v_2^2}
\nonumber\\
m^{'2}_{H44} &=& \frac{v^4(\btto \vt \vth-{a}_u \vo \vth+ \btot \vo \vf-{a}_d \vt \vf)+ 8 \lambda (\vt^2 \vth^2-\vo^2 \vf^2)^2}{v^2 v_1^2 v_2^2}  .
\end{eqnarray*}
\end{small}

The Runge basis helps us see what the lightest Higgs boson mass becomes in the limit that supersymmetry breaking masses are large, $\tilde m\gg M_Z$.  In that case, the $\{ 11\}$ element of the CP even Higgs boson mass matrix is the only diagonal element that stays small. A theorem of linear algebra tells us that the smallest eigenvalue of a positive definite matrix is  smaller than the smallest diagonal element. Therefore, at tree-level we know from this $\{ 11\}$ element the  upper bound on the mass of the lightest CP even Higgs $h$, which is\footnote{Note that such an upper bound would exist even if our  assumption that $b_{ij}$ and $a_i$ are real is not true. For complex $b_{ij}$ and $a_i$ the CP even and pseudoscalar states mix. There are again seven neutral scalars and a Goldstone boson. The 7$\times$7 mass matrix of the physical scalars would again have   $\frac{2\lambda(\sum_i (v_{di}^2-v_{ui}^2))^2}{v^2}$ as the $\{11\}$ element in  the seven dimensional basis in which $V_1$ is the Runge vector. Here the Runge vector in the original eight dimensional space is $V_1^T=\frac{1}{v}(v_1,v_2,v_3,v_4,0,0,0,0)$.}
\bea
m_h^2 &\leq& \frac{2\lambda(\sum_i (v_{di}^2-v_{ui}^2))^2}{v^2}
\nonumber\\
\Rightarrow m_h &\leq& M_Z |\cos^2 \omega \cos 2 \beta_1 +\sin^2 \omega \cos 2 \beta_2|,
\label{eq:treemass}
\eea
where $\tan\beta_i\equiv v_{ui}/v_{di}$ and $\tan \omega \equiv v_2/v_1$.

The leading supersymmetry breaking corrections to this expression are from top squark loops in the same manner as found in the Minimal Supersymmetric Standard Model (MSSM). Thus, the controlling difference between our theory and the MSSM prediction for the Higgs mass is the tree-level expression of \eq{eq:treemass} compared to that of the MSSM, which is 
\beq
m_h\leq M_Z|\cos 2\beta|.
\eeq
One sees that if we set $\tan\beta_1=\tan\beta$, which becomes in both theories the fixed value for the ratio of vevs of the two Higgs doublets that couple to the fermions,  a small additional contribution can be made to the Higgs boson mass in our next generation Higgs theory compared to the MSSM by allowing for larger $\tan\beta_2>\tan\beta_1=\tan\beta$.   

The available gain to the Higgs boson mass in this manner is tiny if $\tan\beta_1\gsim 5$. For lower values of $\tan\beta_1$ the additional Higgs doublet pair contributions can be significant if the mixing angle $\omega$ is larger.  This may be useful since it is a challenge in the MSSM to obtain a Higgs boson mass above the $114\gev$ experimental limit without having too-high superpartner masses that induce fine-tuning in the electroweak sector potential. 

In Fig.~\ref{fig:contour} we plot contours of the tree-level Higgs boson mass in units of $M_Z$ in the plane of $\tan\beta_1$ vs.\ $\sin\omega$, assuming that $\cos 2\beta_2=-1$, which is a good approximation if $\tan\beta_2\gsim 5$.  One sees that as $\sin\omega$ increases, the second Higgs doublet is more responsible for electroweak symmetry breaking and therefore the Higgs mass increases due to larger $\tan\beta_2$. The drawback is that the Higgs bosons that couple to the fermions get smaller vevs, leading to larger Yukawa couplings. The larger top Yukawa coupling could diverge at a low scale. We describe this effect in more detail in  section~\ref{secRG}. In Fig.~\ref{fig:contour} we have also plotted, therefore, the contours of the scale $\Lambda_{SC}$ at which the top Yukawa coupling becomes strongly interacting (i.e., diverges). In Table~\ref{table:higgs mass} we show that with increasing $\sin\omega$ mixing angle, it is possible to have a smaller value of the stop masses so that that the Higgs boson mass is greater than $114\gev$.  The quoted values of $\tilde m_t\equiv \sqrt{\tilde m_{t_1}\tilde m_{t_2}}$ are obtained by assuming that the tree-level contributions are derived from $\tan\beta_1=1.5$ and $\cos 2\beta_2=-1$, and that only the leading order radiative correction is contributing to the Higgs boson mass,
\beq
\Delta m_h^2=\frac{3}{2\pi^2}\frac{m_t^4}{v^2}\log \left(\frac{\tilde m_{t_1}\tilde m_{t_2}}{m_t^2}\right).
\eeq
For higher values of $\sin\omega$, the tree-level contribution increases, thereby putting less pressure on the stop masses to raise the Higgs boson mass above $114\gev$. This is clear from the values in the table, where for higher $\sin\omega$ the needed $\tilde m_t$ values are lower.

\begin{table}[t]
\centering
\begin{tabular}{c c c }
\hline \hline
$\sin\omega$ & $\tilde m_t$ (TeV) & $\Lambda_{SC}$ (GeV) \\
\hline
0 & 2.7 & $2\times 10^{16}$ \\
0.5 & 2.0 & $5\times 10^{8}$ \\
0.7 & 1.4 & $3\times 10^5$ \\
0.9 & 0.8 & $2\times 10^{3}$ \\
\hline
\end{tabular}
\caption{For various values of $\sin\omega$ we show the value of $\tilde m_t$ needed to raise the Higgs boson mass above the experimental limit of $114\gev$ and also the scale $\Lambda_{SC}$ where the top Yukawa coupling diverges. The tree-level contribution is obtained by assuming $\tan\beta_1=1.5$ and $\cos 2\beta_2=-1$.}
\label{table:higgs mass}
\end{table}

We also show in Table~\ref{table:higgs mass} the scale $\Lambda_{SC}$ at which the top Yukawa coupling diverges for the various values of $\sin\omega$.
 As is expected, the larger the value of $\sin\omega$ for some given $\tan\beta_1$, the lower the $\Lambda_{SC}$ scale.  This is the tradeoff one has between a larger Higgs mass prediction and a lower scale of top quark Yukawa coupling divergence. This is reminiscent of the Next-to-Minimal Supersymmetric Standard model (NMSSM) which can have an arbitrarily large Higgs boson mass by adjusting the parameter $\lambda$ in the superpotential interaction $\lambda NH_uH_d$; however, large values of $\lambda$ imply divergences well below a putative unification scale.  This is a common feature in many attempts to solve the Higgs mass bound challenge of supersymmetry.

%%%%%%%%%%%%%%%%%%%%%%%%%%%%%%%%%
\section{Next Generation Higgs boson of Standard Model}

\subsection{Electroweak symmetry breaking and scalar mass matrices}

To add a next generation Higgs boson to the SM is equivalent to postulating a 2HDM with two   scalar doublets $\Phi_1$ and $\Phi_2$ having hypercharge $1/2$. Earlier we discussed the many ways that a second Higgs boson can be added to the spectrum without being incompatible with experiment. There are many options, including type I models and type II models and variants on that theme. The type II structure is most naturally incorporated within supersymmetry using holomorphy as the guiding principle, as we discussed in the previous section. Going beyond that, the most straightforward way to incorporate extra Higgs doublets is to implement a type I structure. In other words, the second Higgs boson (or next full generation) induces no tree-level FCNC by virtue of it having no Yukawa couplings with the SM fermions. 

To ensure no couplings of the second Higgs boson to fermions,  the discrete symmetry,
\begin{equation}
\Phi_2\rightarrow -\Phi_2.
\end{equation}  
can be  imposed, for example. If we allow a soft violation of this symmetry by dimension-two terms we can still avoid tree level FCNC bounds. The most general renormalizable potential for the scalars in which the discrete symmetry is softly broken only by dimension-two terms is,
\bea
V(\Phi_1,\Phi_2) & = & \mu^2_1|\Phi_1|^2+\mu^2|\Phi_2|^2+b (\Phi_1^\dagger \Phi_2 + c.c.)+\lambda_1|\Phi_1|^4+\lambda_2|\Phi_2|^4 \nonumber \\& & +\lambda_3|\Phi_1|^2|\Phi_2|^2
 +\lambda_4 (\Phi_2^\dagger\Phi_1)(\Phi^\dagger_1\Phi_2)+
\left[ \frac{\lambda_5}{2}(\Phi_1^\dagger\Phi_2)^2+c.c.\right].
\label{genpot}
\eea  
The potential above is the potential in the toy model of~\cite{Ambroso:2008kb} once the singlet in their theory gets a vev. Hermiticity requires all the coupling constants in the above potential to be real with the exception of $b$ and $\lambda_5$. A convenient choice of the phase of $\Phi_2$ will allow only one of the two couplings to be real so that the theory is CP violating in general. Note that if both $b$ and $\lambda_5$ vanish the potential is invariant under a global $U(1)$ symmetry for $\Phi_2$. Thus either $b$ or $\lambda_5$ must be non-zero to prevent the pseudoscalar from being a massless Goldstone boson.

Two CP conserving limits of this potential have been considered in the literature~\cite{ferm1}. The limit in which $b=0$ has been called potential A ($V_A$) and the limit in which  $\lambda_5=0$ has been called potential B ($V_B$). Let us now analyze in some detail the electroweak symmetry breaking (EWSB) pattern in these two limits\footnote{For a discussion about the vacuum structure and the possibility of spontaneous violation of CP and electromagnetism in more general 2HDMs see Ref.~\cite{2hdmvac}.}. 

\noindent
\textit{Potential A: $b=0, \lambda_5 \neq 0$}

Potential A can be obtained by strictly imposing the discrete symmetry $\Phi_2\rightarrow -\Phi_2$ which requires $b=0$ in eq.~(\ref{eq:type I potential}),
\bea
V_A(\Phi_1,\Phi_2) & = & \mu^2_1|\Phi_1|^2+\mu^2|\Phi_2|^2+\lambda_1|\Phi_1|^4+\lambda_2|\Phi_2|^4+\lambda_3|\Phi_1|^2|\Phi_2|^2 \nonumber \\
& & +\lambda_4 (\Phi_2^\dagger\Phi_1)(\Phi^\dagger_1\Phi_2)+
\left[ \frac{\lambda_5}{2}(\Phi_1^\dagger\Phi_2)^2+c.c.\right].
\label{eq:type I potential}
\eea
Without loss of generality all the $\lambda_i$ couplings are real. Hermiticity demands it for all $\lambda_i$ except $\lambda_5$, which can always be rotated to real and chosen to be either positive or negative by making $\Phi_2$ absorb its phase.  For definiteness we choose  $\lambda_5 \leq 0$ here. 

The potential must be bounded from below in all field directions. One can test for dangerous runaway directions by parameterizing field excursions arbitrarily large in value.  The following  field directions give us all the unbounded from below (UFB) constraints (see, e.g.,~\cite{UFB}):
\beq
\begin{array}{cc}
\Phi_1^T,\Phi_2^T~{\rm direction} & {\rm UFB~constraint} \\
\hline
(0,a),(0,0)& \lambda_1>0 \\
(0,0),(0,a)& \lambda_2>0 \\
(0,\lambda_2^{1/4}a),(\lambda_1^{1/4}a,0) & \lambda_3+2\sqrt{\lambda_1\lambda_2}>0\\
(0,\lambda_2^{1/4}a),(0,\lambda_1^{1/4}a) & \lambda_3+\lambda_4+\lambda_5+2\sqrt{\lambda_1\lambda_2}>0.
\end{array}
\label{UFB}
\eeq

The most general vacuum expectations values for the two $\Phi_{1,2}$ Higgs fields can be expressed 
as (see, e.g., \cite{DiazCruz:1992uw})
\beq
\Phi_1=\vector{0}{v_1/\sqrt{2}},~~{\rm and}~~\Phi_2=\vector{u_2/\sqrt{2}}{v_2e^{i\xi}/\sqrt{2}}.
\label{vev1}
\eeq
A non-zero $u_2$ would indicate the full breaking of $SU(2)_L\times U(1)_Y$, and in particular the photon would obtain mass. For electromagnetism to remain unbroken when $\Phi_1$ and $\Phi_2$ get vevs the following condition must hold (see for instance ~\cite{DiazCruz:1992uw,Wells:2009kq}),
\beq
\lambda_4 + \lambda_5 <0.
\label{photonmass}
\eeq
The minimization conditions  obtained by setting $dV/d\phi_i=0$ for all real fields $\phi_i$ defined in
\beq
\Phi_1=\vector{\phi_1+i\phi_2}{\phi_3+i\phi_4},~~{\rm and}~~\Phi_2=\vector{\phi_5+i\phi_6}{\phi_7+i\phi_8},
\eeq 
then ensure that EWSB is proper and the doublets get the  vevs, 
\beq
\Phi_1=\vector{0}{v_1/\sqrt{2}},~~{\rm and}~~\Phi_2=\vector{0}{v_2/\sqrt{2}}
\label{vev2}
\eeq
where $v_1\geq0$ and we can always choose $v_2\geq0$  by a convenient choice of sign of the doublet field $\Phi_2$. The minimization conditions $dV/d\phi_3=0$ and $dV/d\phi_7=0$ given by~\cite{DiazCruz:1992uw,Wells:2009kq},
\beq
\begin{array}{cl}
\phi_3: & \mu^2_1+\frac{\lambda_3+\lambda_4+\lambda_5}{2}v^2_2+\lambda_1v^2_1=0 \\
\phi_7: & \mu^2_2+\frac{\lambda_3+\lambda_4+\lambda_5}{2}v^2_1+\lambda_2v^2_2=0
\end{array}
\label{minim}
\eeq    
can be used to eliminate the parameters $\mu_1$ and $\mu_2$~\cite{Barroso:2007rr}.
 
We need to check if this solution is stable. To do that we require that the second derivative of the potential, i.e.\ the mass matrix, be positive definite.  We can find the mass eigenvalues are by solving four $2\times 2$ matrices. These matrices arise from $\phi_k\phi_{k+4}$ mixing for $k=1,2,3,4$. We simplify the entries in these matrices by substituting $\mu_1^2$ and $\mu_2^2$ from \eq{minim}. To begin with, we look at the $\phi_1\phi_5$ and $\phi_2\phi_6$ mixings, which have the same $2\times 2$ mass matrix:
\beq
{\cal M}^2_{\phi_1\phi_5}={\cal M}^2_{\phi_2\phi_6}=
\left( \begin{array}{cc}
 -\frac{\lambda_4+\lambda_5}{2}v^2_2 & \frac{\lambda_4+\lambda_5}{2}v_1v_2 \\
 \frac{\lambda_4+\lambda_5}{2}v_1v_2 & -\frac{\lambda_4+\lambda_5}{2}v^2_1
 \end{array}\right)
 \eeq
 which leads to four eigenstates
 \bea
 m^2_{G^\pm} & = & 0~~{\rm (charged~Goldstone~bosons)} \\
 m^2_{H^\pm}& =& -\frac{\lambda_4+\lambda_5}{2}(v^2_1+v^2_2)~~{\rm (charged~Higgs~bosons)}.
\eea
The mixing angle is,
\beq
\tan \omega = \frac{v_2}{v_1}.
\eeq
Now let us look at $\phi_4\phi_8$ mixing:
\bea
{\cal M}^2_{\phi_4\phi_8}=
\left( \begin{array}{cc}
 -\lambda_5v^2_2 & \lambda_5 v_1v_2  \\
\lambda_5 v_1v_2 & -\lambda_5 v^2_1
 \end{array}\right).
 \eea
This leads to two eigenstates
\bea
m^2_{G} & = & 0 ~~{\rm (neutral~Goldstone~bosons)}  \\
m^2_{A}&=& -\lambda_5(v^2_1+v^2_2)~~{\rm (neutral~pseudoscalar~boson)}.
\eea
The mixing angle is again $\tan \omega = v_2/v_1$. Finally, there is $\phi_3\phi_7$ mixing:
\beq
{\cal M}^2_{\phi_3\phi_7}=
\left( \begin{array}{cc}
2\lambda_1v^2_1 & (\lambda_3+\lambda_4+\lambda_5)v_1v_2 \\
(\lambda_3+\lambda_4+\lambda_5)v_1v_2 & 2\lambda_2v^2_2
\end{array}\right).
\eeq
This is the $2\times 2$ mass matrix for the two physical neutral scalar Higgs bosons of the theory, $h$ and $H$. 
The mixing angle to rotate from $\{\phi_3,\phi_7\}$ basis to $\{H,h\}$ basis is usually called $\alpha$, which is defined by convention to satisfy
\beq
\vector{H}{h}=
\left( \begin{array}{cc}
\cos\alpha & \sin\alpha \\
-\sin\alpha & \cos\alpha
\end{array}\right)\vector{\phi_3}{\phi_7}.
\eeq
\begin{figure}[t]
\begin{center}
\includegraphics[width=\columnwidth]{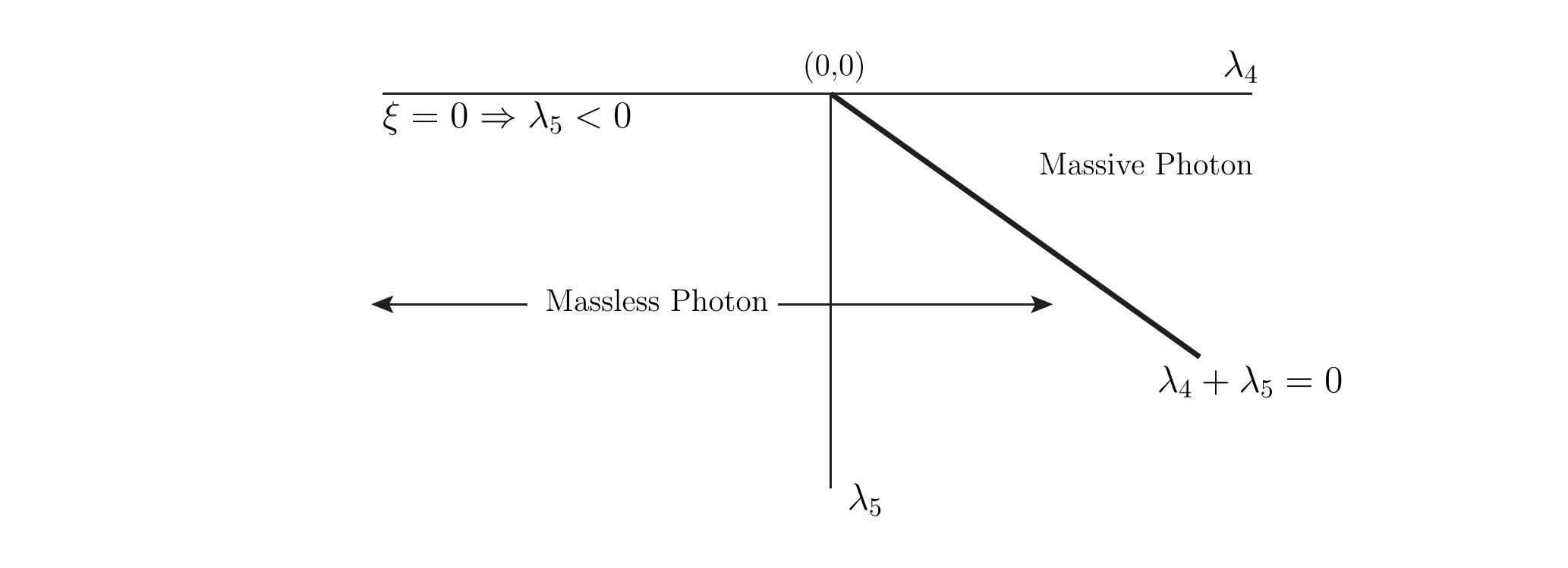}
\end{center}
\caption{Parameter space for massive and massless photon in a type I 2HDM with potential A.}
\label{fig:phaseA}
\end{figure}
The solutions are obtained by simple eigenvalue, eigenvector analysis of the $2\times 2$ matrix, and one obtains
\begin{equation}
\tan 2\alpha =\frac{ \lambda_{345} v_1 v_2}{\lambda_1 v_{1}^2 - \lambda_2 v_2^2}
\end{equation}
where $ \lambda_{345}=\lambda_3+\lambda_4+\lambda_5$. We will use the convention that $H$ is the `SM-like' Higgs and hence we will always take $\alpha<\pi/4$. The masses of the CP-even scalars are,
\begin{equation}
 m^{2}_{h,H}    =  \lambda_1 v_1^2 + \lambda_2 v_{2}^2 
            \pm \sqrt{ (\lambda_1 v_1^2 - \lambda_2 v_{2}^2)^2 +  \lambda_{345} v_1 v_2^2} .
\end{equation}.

We  require that the mass matrix be positive definite, which puts important constraints on the parameters of the theory. For example, from the charged Higgs and pseudo-scalar Higgs boson masses we know that
\beq
\lambda_4+\lambda_5<0,~~{\rm and}~~\lambda_5<0
\eeq
is required. Note that the first condition is the same as \eq{photonmass}, the condition that ensures that the photon remains massless.  Fig.~\ref{fig:phaseA} plots the parameter space in the $\lambda_4$ vs.\ $\lambda_5$ plane that corresponds to massive photon and massless photon cases in agreement with Ref.~\cite{DiazCruz:1992uw}. 

\noindent
\textit{Potential B: $\lambda_5=0,~b\neq0$}

Potential B can be obtained by imposing a global $U(1)$ symmetry for $\Phi_2$ and allowing it to be broken only softly by dimension-two terms like $b\Phi^\dagger_1\Phi_2$,
\bea
V(\Phi_1,\Phi_2) & = & \mu^2_1|\Phi_1|^2+\mu^2|\Phi_2|^2+b(\Phi^\dagger_1\Phi_2+c.c.) +\lambda_1|\Phi_1|^4+\lambda_2|\Phi_2|^4+\lambda_3|\Phi_1|^2|\Phi_2|^2 \nonumber \\
& & +\lambda_4 (\Phi_2^\dagger\Phi_1)(\Phi^\dagger_1\Phi_2).
\label{potb}
\eea
Without loss of generality all the $\lambda_i$ couplings are real due to hermiticity. The coupling $b$ can be rotated to real by $\Phi_2$ absorbing its phase. 

The conditions for the potential to be bounded from below in all field directions are the same as \eq{UFB} with $\lambda_5=0$ as these conditions are determined by the quartic couplings so that the additional bilinear term does not affect them. The most general vacuum expectations values for the two $\Phi_{1,2}$ Higgs fields can again be expressed by \eq{vev1}.

The minimization condition $dV/d\phi_1=0$~\cite{DiazCruz:1992uw} in this case is given by,
\beq
 bu_2+\frac{\lambda_4}{2} u_2v_1v_2\cos\xi=0 
\label{eq:minimize}
\eeq
It is clear that for
\beq
b+\frac{\lambda_4}{2} v_1v_2\cos\xi\neq0
\eeq
it is required that $u_2=0$ from the $\phi_1$ minimization condition in \eq{eq:minimize}. This ensures that electromagnetism is not broken and the photon remains massless as the doublets get vevs. This condition and the other minimization conditions then ensure that the doublets get vevs of the form, 
\beq
\Phi_1=\vector{0}{v_1/\sqrt{2}},~~{\rm and}~~\Phi_2=\vector{0}{v_2/\sqrt{2}}
\label{vev}
\eeq
where $v_1\geq0$ and we can always choose $v_2\geq0$  by a convenient choice of sign of the doublet field $\Phi_2$, which can then be absorbed in a redefinition of the coupling $b$. The minimization conditions $dV/d\phi_3=0$ and $dV/d\phi_7=0$ given by~\cite{DiazCruz:1992uw},
\beq
\begin{array}{cl}
\phi_3: & \mu^2_1+b v_2 +\frac{\lambda_3+\lambda_4}{2}v^2_2+\lambda_1v^2_1=0 \\
\phi_7: & \mu^2_2+b v_1+\frac{\lambda_3+\lambda_4}{2}v^2_1+\lambda_2v^2_2=0
\end{array}
\label{minim2}
\eeq    
can be used to eliminate the parameters $\mu_1$ and $\mu_2$~\cite{Barroso:2007rr}. 

\begin{figure}[t]
\begin{center}
\includegraphics[width=\columnwidth]{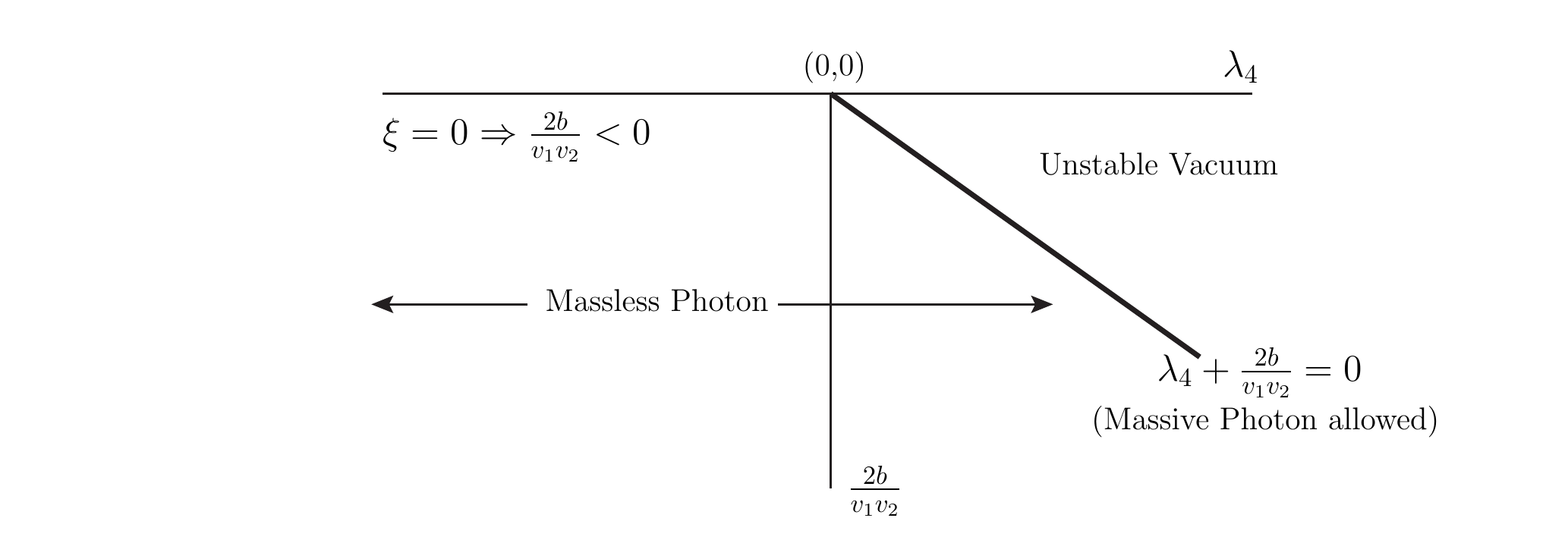}
\end{center}
\caption{Parameter space for the different patterns of EWSB in the type I 2HDM with potential B.}
\label{fig:phaseB}
\end{figure}

Let us now look at the mass matrix after substituting $\mu_1^2$ and $\mu_2^2$ from \eq{minim2}. Let us first look at the $\phi_1\phi_5$ and $\phi_2\phi_6$ mixings, which have the same $2\times 2$ mass matrix:
\beq
{\cal M}^2_{\phi_1\phi_5}={\cal M}^2_{\phi_2\phi_6}=
\left( \begin{array}{cc}
 -\frac{\lambda_4}{2} v^2_2-b v_2/v_1 & \frac{\lambda_4}{2} v_1v_2+b \\
 \frac{\lambda_4}{2} v_1v_2+b & -\frac{\lambda_4}{2}v^2_1-b v_1/v_2
 \end{array}\right)
 \eeq
 which leads to four eigenstates
 \bea
 m^2_{G^\pm} & = & 0~~{\rm (charged~Goldstone~bosons)} \\
 m^2_{H^\pm}& =& -\frac{v^2_1+v^2_2}{2}\left(\lambda_4+\frac{2b}{v_1 v_2}\right)~~{\rm (charged~Higgs~bosons)},
\eea
the mixing angle being,
\beq
\tan \omega=\frac{v_2}{v_1}.
\eeq
Now let us look at $\phi_4\phi_8$ mixing:
\bea
{\cal M}^2_{\phi_4\phi_8}=
\left( \begin{array}{cc}
 -b v_2/v_1 & b  \\
b & -b v_1/v_2
 \end{array}\right).
 \eea
This leads to two eigenstates
\bea
m^2_{G^0} & = & 0 ~~{\rm (neutral~Goldstone~boson)}  \\
m^2_{A}&=& -\frac{b}{v_1 v_2}(v_1^2+v_2^2)~~{\rm (neutral~pseudoscalar~boson)},
\label{pseudoB}
\eea
where the mixing angle is again given by $\tan \omega= v_2/v_1$. Finally, there is $\phi_3\phi_7$ mixing:
\beq
{\cal M}^2_{\phi_3\phi_7}=
\left( \begin{array}{cc}
2\lambda_1 v^2-b v_2/v_1 & (\lambda_3+\lambda_4)v_1 v_2+b  \\
(\lambda_3+\lambda_4)v_1 v_2+b & 2\lambda_2v^2_2-b v_1/v_2
\end{array}\right).
\eeq
We have again,
\beq
\vector{H}{h}=
\left( \begin{array}{cc}
\cos\alpha & \sin\alpha \\
-\sin\alpha & \cos\alpha
\end{array}\right)\vector{\phi_3}{\phi_7}.
\eeq
The masses and mixing angle are,
\begin{equation}
\tan 2\alpha =\frac{  \lambda_{34} v_1 v_2+b}{\lambda_1 v_{1}^2 - \lambda_2 v_2^2+b\frac{v^2_1-v^2_2}{2 v_1 v_2}}
\end{equation}
where $  \lambda_{34}=\lambda_3+\lambda_4$, and
\begin{equation}
 m^{2}_{h,H}    =  \lambda_1 v_1^2 + \lambda_2 v_{2}^2 -b\frac{v^2_1+v^2_2}{2 v_1 v_2}
            \pm \sqrt{ \left(\lambda_1 v_1^2 - \lambda_2 v_{2}^2+b\frac{v^2_1-v^2_2}{2 v_1 v_2}\right)^2 + ( \lambda_{34} v_1 v_2+b)^2}.
            \label{hmassB}
\end{equation}
Requiring the charged and pseudoscalar Higgs mass to be positive gives us,
\beq
\lambda_4+\frac{2b}{v_1 v_2}<0~~~~~\frac{2b}{v_1 v_2}<0.
\eeq
Fig.~\ref{fig:phaseB} shows the the regions in the parameter space which lead to the different  patterns of EWSB for potential B. 

An analysis of the general potential in \eq{genpot} is complicated by the fact that both $b$ and $\lambda_5$ cannot be chosen to be real by redefining the phase of $\Phi_2$. This leads to a mixing between the pseudoscalar and CP even states. If both $b$ and $\lambda_5$ are assumed to be real, however, the masses and mixing angles for the general potential in \eq{genpot}, assuming that the vacuum is invariant under CP and electromagnetism, are,
\beq
m^2_{H^\pm} = -\frac{v^2_1+v^2_2}{2}\left(\lambda_4+\lambda_5+\frac{2b}{v_1 v_2}\right)
\eeq
\beq
m^2_{A}= -(v_1^2+v_2^2)\left(\lambda_5+\frac{b}{v_1 v_2}\right)
\label{genma}
\eeq
\beq
 m^{2}_{h,H}    =  \lambda_1 v_1^2 + \lambda_2 v_{2}^2 - b\frac{v^2_1+v^2_2}{2 v_1 v_2}
            \pm \sqrt{ \left(\lambda_1 v_1^2 - \lambda_2 v_{2}^2+b\frac{v^2_1-v^2_2}{2 v_1 v_2}\right)^2 +  \lambda_{345} v_1 v_2^2}
            \eeq
\beq
                        \tan \omega=\frac{v_2}{v_1}
\eeq
\begin{equation}
\tan 2\alpha =\frac{ \lambda_{345} v_1 v_2+b}{\lambda_1 v_{1}^2 - \lambda_2 v_2^2+b\frac{v^2_1-v^2_2}{2 v_1 v_2}}
\label{tanalpha}
\end{equation}
where $ \lambda_{345}=\lambda_3+\lambda_4+\lambda_5.$ Thus in 2HDMs from the original eight degrees of freedom we get three Goldstone bosons and five physical scalars. The Goldstone bosons are absorbed as longitudinal modes by $W^{+/-}$ and $Z$. The tree level masses of the $W^{+/-}$ and $Z$ are,
\begin{equation}
M_W = M_Z \cos \theta_W= \frac{g}{2}\sqrt{v_1^2 +v_2^{2}}.
\end{equation}

%\subsection{Experimental Constraints on the Type I Two-Higgs Doublet Model}

We now give some of the constraints that this model experiences when requiring compatibility with all past experiment. 

\subsection{Indirect Constraints}
\textit{Constraints due to virtual  $H^{+/-}$ effects}

\begin{figure}[t]
\begin{center} 
\includegraphics[width=0.7\columnwidth]{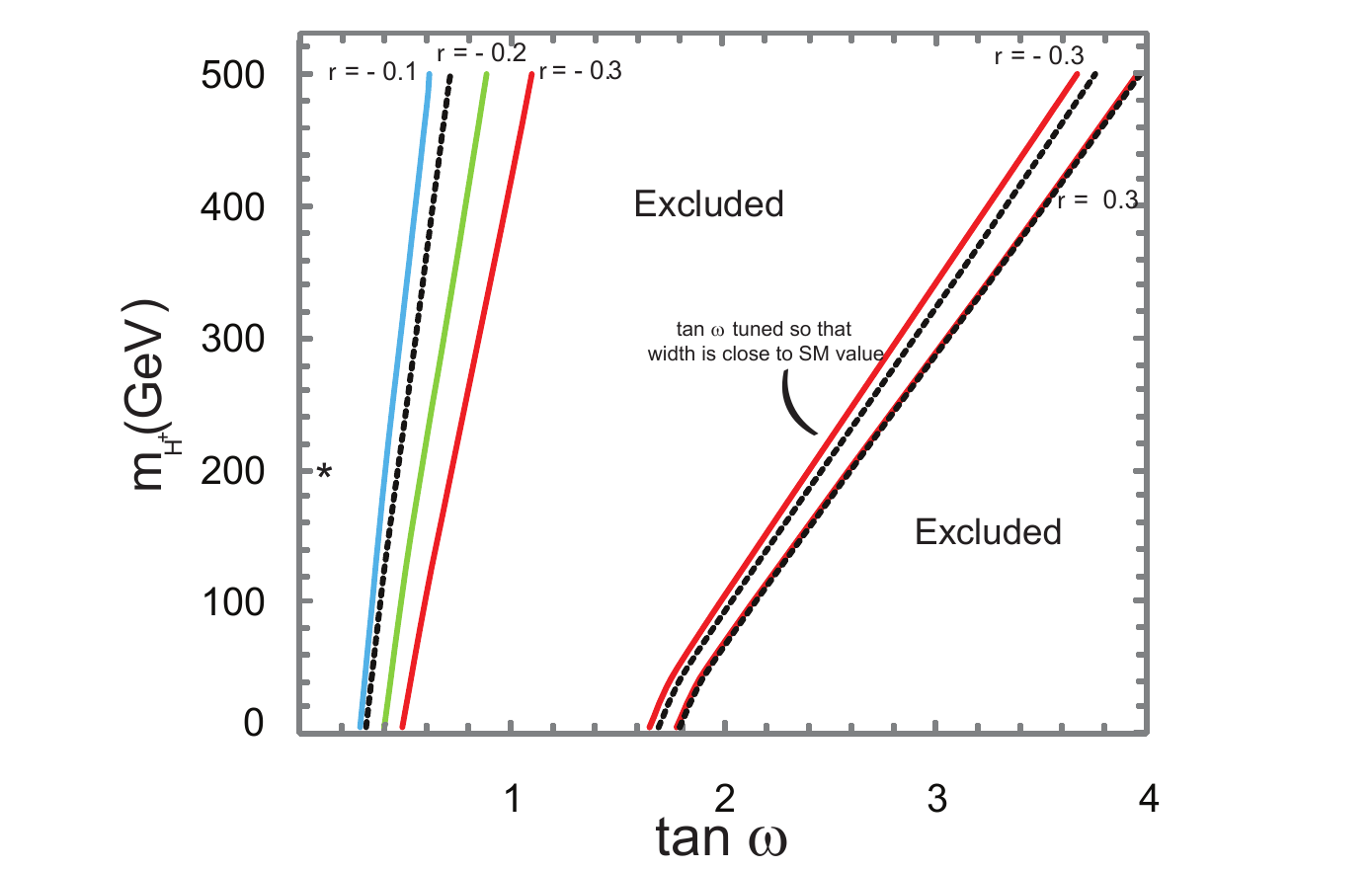}
\end{center}
\caption{The solid lines show contours of constant $r$, the fractional deviation of the type I 2HDM value of the $b\rightarrow s\gamma$ width from the SM value.  The dashed lines are the boundaries of the region in the $m_{H^+}$-$\tan \omega$ plane excluded by constraints from the  $b\rightarrow s\gamma$ branching rate at two standard deviations. As one can see from \eq{eq:bsgamma} there are two different ways of satisfying the constraint: (1) if $\tan \omega$ is small the type I 2HDM width is close to the SM value, or (2) $\tan^2 \omega$ can be tuned to a higher value so that $\left|\sum_{i=c,t}\lambda_i C_{7i}(m_b) \right| $ in \eq{eq:bsgamma} is close to the SM value ($C_{7i}\simeq -C_{7i}^{SM}$). Thus there are two disconnected allowed regions in Fig.~\ref{bsgamma}. The star shows our choice of $m_{H^+}$ and $\tan \omega$ that we use for collider simulations later.}
\label{bsgamma}
\end{figure}

In  the type I 2HDM the coupling of $H^{+/-}$ to  the fermions is proportional to $\tan {\omega}$. Thus in  the type I 2HDM, unlike the type II 2HDM, all the constraints coming from processes involving virtual $H^{+/-}$ can be met for small enough $\tan{\omega}$. The strongest such constraint comes from the  $b\rightarrow s\gamma$ branching rate. An expression for the width of the process in type I 2HDMs is~\cite{hew},
\begin{eqnarray}
\Gamma_{b\rightarrow s\gamma}&=&\frac{\alpha G_F^2 m_b^2}{128 \pi^4}\left|\sum_{i=c,t}\lambda_i C_{7i}(m_b) \right| ^2 \nonumber\\
C_{7i}(m_b)&=&\eta^{-16/23}[C_{7i}(M_W)-\frac{116}{135}(\eta^{10/23}-1)
-\frac{116}{378}(\eta^{28/23}-1)]\nonumber\\
C_{7i}(M_W)&=&A_{W}(m_i^2/M_W^2) + \tan^2\omega~ A_{Hi}(m_i^2/m_{H^+}^2).
\label{eq:bsgamma}
\end{eqnarray}
Here $\lambda_i=V_{is}^* V_{ib}$, $\eta=\alpha_s(m_b)/\alpha_s(M_W)$, $A_W$ is the SM contribution due to $W^{+/-}$ loops and $A_H$ is the additional contribution due to triangle diagrams involving the charged Higgs. For expressions of $A_W$ and $A_H$ see for instance~\cite{hew}. The SM value for the width can be obtained by simply putting $\omega=0$ or taking the limit $m_{H^+}\rightarrow \infty$ in the above equation. 

Let the fractional deviation of the type I 2HDM value from the SM value be,
\beq
r=\frac{\Gamma^{2HDM-I}_{b\rightarrow s\gamma}-\Gamma^{SM}_{b\rightarrow s\gamma}}{\Gamma_{b\rightarrow s\gamma}^{SM}}.
\eeq
Fig.~\ref{bsgamma} shows the contours of constant $r$  in the $m_{H^+}$-$\tan \omega$ plane. The dashed lines  in  Fig.~\ref{bsgamma} show the region excluded by experiments at 2$\sigma$ level. To obtain the dashed curves we have used the world average for the experimental value of the branching fraction~\cite{hfag},
\begin{displaymath}
B(\bar{B}\rightarrow X_s \gamma)=(3.55 \pm 0.24^{+0.09}_{-0.10} \pm 0.03)\times 10^{-4}.
\end{displaymath}
We have rescaled the theoretical value obtained using \eq{eq:bsgamma} to reproduce the NNLO SM prediction~\cite{nnlo},
\begin{displaymath}
B(\bar{B}\rightarrow X_s \gamma)=(3.15 \pm 0.23)\times 10^{-4}
\end{displaymath}
in the $m_{H^+}\rightarrow \infty$ limit and added the error associated with this value in quadrature to the experimental error. From \eq{eq:bsgamma} one can see that there are two different ways of satisfying the constraint: (1) if $\tan \omega$ is small the type I 2HDM width is close to the SM value, or (2) $\tan^2 \omega$ can be tuned to a higher value so that $\left|\sum_{i=c,t}\lambda_i C_{7i}(m_b) \right| $ in \eq{eq:bsgamma} is close to its SM value ($C_{7i}\simeq -C_{7i}^{SM}$). Thus there are two disconnected allowed regions in Fig.~\ref{bsgamma}. We can see from Fig.~\ref{bsgamma} that the constraint is satisfied for all values of $m_{H^+}$ if,
\begin{equation}
\tan \omega < 0.32 
\end{equation}
which translates to,
\begin{equation}
v_2< 75 ~{\rm GeV}.
\end{equation}
For $m_{H^+}=200$ GeV we obtain the constraint $\tan \omega < 0.47~ (v_2< 105$ GeV). Another constraint due to virtual  $H^{+/-}$ effects comes from  $B_d^0-\bar{B}_d^0$ oscillations. This, however,  puts a weaker constraint than the $b\rightarrow s \gamma$ process at the $2\sigma$ level~\cite{maltoni}. 

It is important to note that a small $v_{2}$ not only implies a small $\omega$ but also suggests a small $\alpha$ for  a 2HDM with the general potential in \eq{genpot}. This can be seen from the expression for $\tan 2 \alpha$  in \eq{tanalpha}. Using \eq{genma} we can obtain the expression for $\tan 2\alpha$ for $\tan\omega=v_2/v_1\ll 1$, 
\beq
\tan 2\alpha \approx\frac{\lambda_3+\lambda_4-(m_A^2/v_1^2) }{\lambda_1-\frac{1}{2}(m_A^2/v_1^2+\lambda_5)}\tan{\omega}
\eeq
which  shows that  a small $v_2$ suggests that $\alpha$ should be small too. \label{smallv2}

\noindent
\textit{The $\rho$ parameter}

In the 2HDM the $\rho$-parameter gets additional contributions from corrections to $M_W$ and $M_Z$ due to loops of the scalars. With respect to the SM theory with a Higgs boson mass of $m_h=120\gev$, the range of $\Delta\rho$ that can be tolerated~\cite{pdg} by replacing the single SM Higgs boson with the 2HDM is $0.0000\lsim \Delta\rho\lsim 0.0012$ at the 68\% CL, where $\Delta\rho=\rho(2HDM)-\rho(m_h^{SM})$. The 2HDM computation needed for this can be found in Ref.~\cite{ber}. One finds that it is easy to satisfy precision electroweak constraints for a 2HDM with masses in the neighborhood of $50-300\gev$, as we shall see later in our example benchmark points that we use to study collider signatures. 

\subsection{Collider Constraints}

%\noindent
%\textit{Collider Constraints}

\begin{figure}[t]
\begin{center} 
\includegraphics[width=0.8\columnwidth]{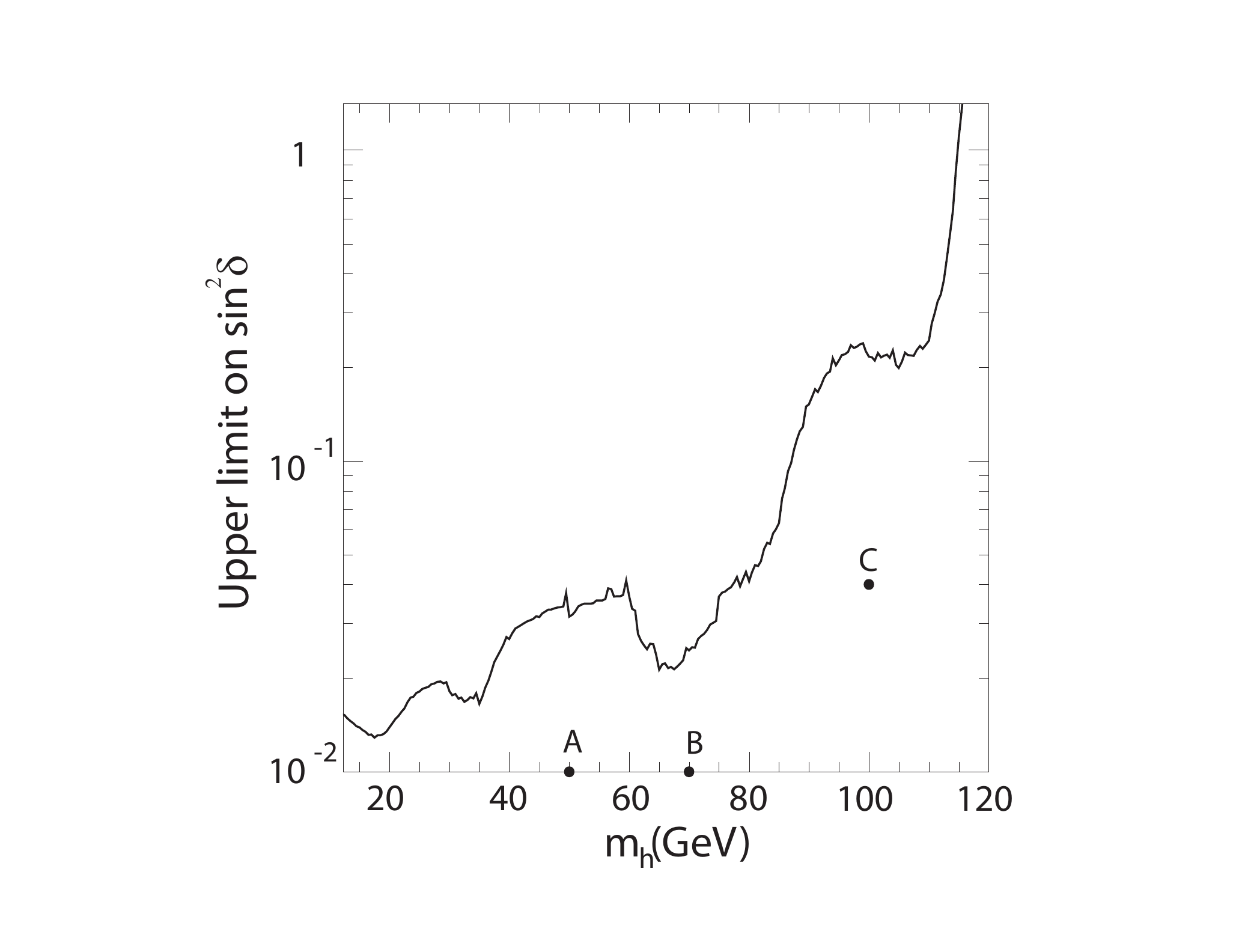}
\end{center}
\caption{Upper bounds on $\sin^2 \delta$ as a function of $m_h$ derived from searches for the Higgsstrahlung process $e^+e^- \rightarrow Zh$ at LEP with center of mass energy $\sqrt{s}$ = 91-209 GeV.  It has been assumed that $h$ decays entirely to $bb$. We also plot the three parameter sets A, B and C that we have chosen for simulations (Table~\ref{tab:input}) on the $\sin^2 \delta$ - $m_h$ plane. The curve has been reproduced from Ref.~\cite{c}.}
\label{sin2delta}
\end{figure}

The fact that none of the scalars $h,H$ or $A$ were seen at LEP puts constraints on their masses $m_h, m_H, m_A$ and $\delta=|\alpha-\omega|$. We are especially interested in the cases when the decay modes $A\rightarrow Zh$ or $ H\rightarrow hh/AA$ are kinematically allowed, i.e.\ when the conditions $|m_A-m_h|>m_Z$ or $m_H > 2 m_{h/A}$  hold respectively.\label{collider}

The major constraint on the mass of the non SM-like CP-even Higgs, $m_h$ comes from  the non-discovery of  $h$ produced by the Higgstrahlung process $e^+  e^- \rightarrow hZ$ at LEP. The cross section of  the Higgstrahlung process in the 2HDM is suppressed by a factor of $\sin^2 \delta$ with respect to the cross section of the same process in the SM. It is therefore possible to have  $h$ as light as we wish by choosing a sufficiently small value of $\delta$.  Upper bounds on the cross section for this process for a particular value of $m_h$ therefore give upper bounds on  $\sin^2 \delta$. Fig.~\ref{sin2delta}  shows the LEP2 upper bounds~\cite{c} on $\sin^2 \delta$ vs $m_h$ assuming $h$ decays entirely to $b\bar{b}$. As one can see from the figure there are no constraints on $m_h$ at all, if $\sin \delta <0.1$, and no constraints on $\sin \delta$ if $m_h>114 $ GeV. If we take $\sin \delta <0.2$, $m_h\approx 100$ GeV is safely within allowed limits. 

Another important process that could have been potentially seen at LEP is the associated production process $e^+ e^- \rightarrow hA$ . The cross section for this process is proportional to $\cos^2 \delta$ in the 2HDM. Analysis of LEP2 data, assuming $|m_A-m_h|>m_Z$ so that $A$ predominantly decays as $A\rightarrow Zh$, has been done  by the DELPHI collaboration~\cite{del} and the process has been found to be unconstrained. Even if this condition does not hold there are no constraints on $\delta$ if we have $m_A +m_h > 200$ GeV~\cite{c}.

The constraints on the pseudoscalar mass $m_A$ are much weaker. The cross section of associated production process, $e^+  e^- \rightarrow HA$,  is proportional to $\sin^2 \delta$. The results of LEP analyses motivated by this process put upper limits on $\sin^2 \delta$ for a given $m_A$ and $m_H$ (see Ref.~\cite{c} for details). If the dominant decay mode is $H\rightarrow AA$ there  are no constraints at all if $m_H>120$ GeV and $m_A>50$ GeV~\cite{c}.

As far as the charged Higgs is concerned, the LEP2 direct search constraints from the process $e^+ e^- \rightarrow Z^{*}\rightarrow H^+H^-$ place a lower limit of  76.7 GeV on $m_H^+$~\cite{abdal}. The Tevatron search for $t\rightarrow H^+ b$ puts a 95$\%$ CL upper bound on $B(t\rightarrow H^+ b)$  at $0.1-0.3$ in the mass range $90$ GeV$<m_{H^+}<150$ GeV assuming B($H^+\rightarrow c\bar{s}$) to be 100$\%$~\cite{charged}. For almost the entire mass range  $B(t\rightarrow H^+ b)<0.1$ is allowed.  In the type I 2HDM this branching ratio is proportional to $\tan^2 \omega$, so that we obtain  $\tan \omega~(\approx s_{\omega}) < 0.3 $ for $m_{H^+}=90$ GeV and  $\tan \omega < 0.8$ $(s_{\omega}<0.6) $  for $m_{H^+}=150$ GeV by requiring $B(t\rightarrow H^+ b)<0.1$. The limit on $\tan\omega$ is actually too conservative for $m_{H^+}=90$ GeV  as the upper bound on the branching fraction for this mass is about 0.3. 

Finally let us consider the SM-like Higgs boson $H$. If $m_H< 2m_{h/A}$, $H$ would predominantly decay into quarks with branching ratios very similar to that in SM. In this case the process $e^+  e^- \rightarrow HZ$ puts  a weaker lower bound on $m_H$ in  2HDMs  than  the SM value $114$ GeV because the  2HDM cross section of this process  is smaller than the SM cross section by a factor $\cos^2 \delta$. If $ H\rightarrow hh/AA$ is allowed there are of course no constraints on $\delta$  if $m_H>115$ GeV, and for  $\delta> 0$, $m_H$ can be even smaller~\cite{c}. We will take $m_H=120$ GeV for all the parameter sets we use in our simulations in section~\ref{secColl}.

%%%%%%%%%%%%%%%%%%%%%%%%%%%%%%%%%%%%%
\section{Yukawa Coupling Perturbativity}
\label{secRG}

If there exists at least one Higgs boson with a vev that does not couple to a fermion $f$, the Yukawa coupling of that fermion $\lambda_f$ must necessarily be greater than its corresponding would-be SM value.  Larger Yukawa couplings in the theory run the risk of renormalizing to strong coupling at a lower scale than desired. In our study this is a consideration that must be explored, since our emphasis is on next generation Higgs bosons that do not couple to the fermions. Therefore, in this section we quantify where Yukawa couplings blow up in renormalization group evolution as a function of the vev of fermiophobic Higgs doublets in the theory. The effect is particularly pronounced for the top quark Yukawa coupling, since it is associated with the highest mass fermion in the theory.

The one loop Renormalization Group Equations (RGEs) for the Yukawa ($\lambda_f$)~\cite{Barger:1992ac} and  gauge couplings (see for eg.~\cite{Grimus:2004yh})  in an extension of the SM with $n_d$ extra Higgs doublets is,
\bea
\frac{d \lambda_t}{d (\ln \mu)}&=&\frac{\lambda_t}{16 \pi^2}\left( \frac{3}{2}\lambda_t^2 -\frac{3}{2} \sum_D |V_{tD}|^2 \lambda_D^2+ S -8 g_3^2 -\frac{9}{4} g_2^2- \frac{17}{12} g'^2\right)
\label{RGEU}
\\
\frac{d \lambda_b}{d (\ln \mu)}&=&\frac{\lambda_b}{16 \pi^2}\left( \frac{3}{2}\lambda_b^2 -\frac{3}{2} \sum_U |V_{Ub}|^2 \lambda_U^2+ S -8 g_3^2 -\frac{9}{4} g_2^2- \frac{5}{12} g'^2\right)
\label{RGED}
\\
\frac{d \lambda_\tau}{d (\ln \mu)}&=&\frac{\lambda_\tau}{16 \pi^2}\left(\frac{3}{2}\lambda_\tau^2 +S-\frac{9}{4} g_2^2- \frac{15}{4} g'^{2}\right)
\label{RGEL}
\\
\frac{d g'}{d (\ln \mu)}&=&\frac{\frac{41+n_d}{6} g'^3}{16 \pi^2},~~~~\frac{d g_2}{d (\ln \mu)} =\frac{\frac{-19+n_d}{6} g_2^3}{16 \pi^2},~~~~\frac{d g_3}{d (\ln \mu)} =-\frac{7 g_3^3}{16 \pi^2},
\eea
where $S=3\sum_U  \lambda_U^2+3 \sum_D  \lambda_D^2+ \sum_E  \lambda_E^2$. $U,D$ and $E$ denote the up-type quarks, the down-type quarks and the leptons while $g'$, $g_2$ and $g_3$ are the $U(1)_Y$, $SU(2)_L$ and $SU(3)$ gauge couplings. Note that the summation over $U, D$ does not include a summation over colors. There are similar equations for the other up-type quarks ($t\rightarrow U$ in \eq{RGEU}), the other down-type quarks ($b\rightarrow D$ in \eq{RGED} and the other leptons ($\tau\rightarrow E$ in \eq{RGEL}). For the type I 2HDM $n_d=1$. 
\begin{figure}[t]
\centering
\includegraphics[width=0.7\columnwidth]{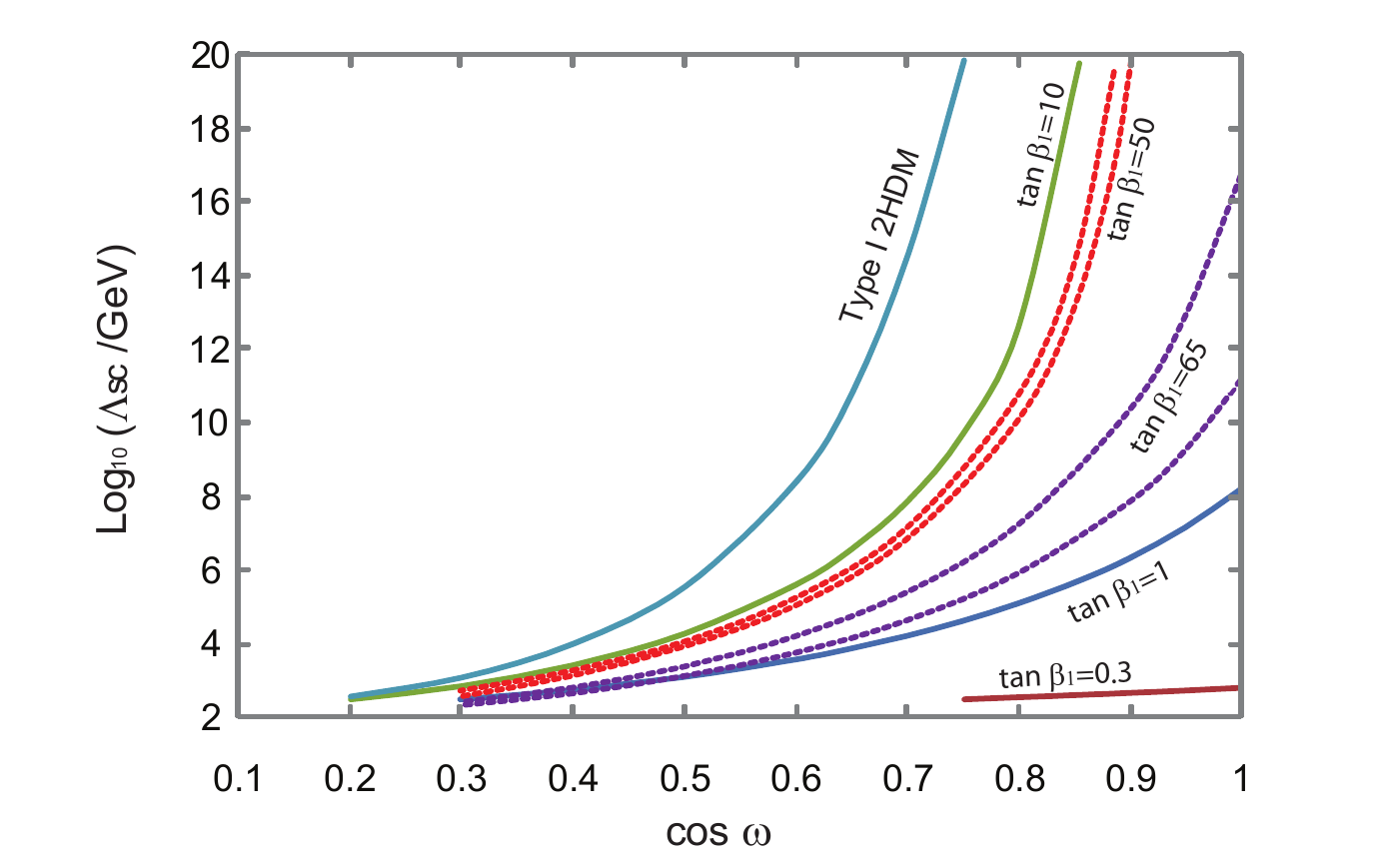}
\caption{$\Lambda_{SC}$, the energy scale at which either one of the three large Yukawa couplings $\lambda_t, \lambda_b$ or $\lambda_\tau$ becomes larger than $4\pi$, as a function of $c_{\omega}$. We show curves for the type I 2HDM case  as well as the the SUSY 2HGM with different values of $\tan {\beta_1}$. For the $\tan {\beta_1}=50$ and $\tan {\beta_1}=65$ cases we show two dashed curves corresponding to the minimum (left) and maximum (right) allowed value of $m_b(M_Z)$ in \eq{eq:mb}. For the mean value  $m_b(M_Z)=2.85$ GeV in \eq{eq:mb}, for $\tan {\beta_1} \geq55$ it is the bottom Yukawa which becomes strong ($>4 \pi$). We have used the Renormalization Group Equations at one loop level.}
\label{RGflow}
\end{figure}

For supersymmetric models with $n_g$ extra generations of Higgs doublet pairs than MSSM the one loop RGEs  for Yukawa~\cite{Barger:1992ac} and gauge couplings (see for eg.~\cite{Kolda:1995fg}) are,
\bea
\frac{d \lambda_t}{d (\ln \mu)}&=&\frac{\lambda_t}{16 \pi^2}\left( 3\lambda_t^2 + \sum_D |V_{tD}|^2 \lambda_D^2+ 3 \sum_U  \lambda_U^2-\frac{16}{3} g_3^2 -3 g_2^2- \frac{13}{9} g'^2\right)\label{RGSU}\\
\frac{d \lambda_b}{d (\ln \mu)}&=&\frac{\lambda_b}{16 \pi^2}\left(3\lambda_b^2 + \sum_U |V_{Ub}|^2 \lambda_U^2+ 3 \sum_D  \lambda_D^2+\sum_E  \lambda_E^2-\frac{16}{3} g_3^2 -3 g_2^2- \frac{7}{9} g'^2\right)\label{RGSD}\\
\frac{d \lambda_\tau}{d (\ln \mu)}&=&\frac{\lambda_\tau}{16 \pi^2}\left(3\lambda_\tau^2 + 3\sum_D  \lambda_D^2+  \sum_E  \lambda_E^2 -3 g_2^2- 3 g'^2\right)\label{RGSL}\\
\frac{d g'}{d (\ln \mu)}&=&\frac{(11+n_g) g'^3}{16 \pi^2},~~~~\frac{d g_2}{d (\ln \mu)} =\frac{(1+n_g) g_2^3}{16 \pi^2},~~~~\frac{d g_3}{d (\ln \mu)} =-\frac{3 g^3}{16 \pi^2}.
\eea
For the supersymmetric two Higgs generation Model (SUSY 2HGM) of section~\ref{susy} $n_g=1$. There are similar equations for the other up-type quarks ($t\rightarrow U$ in \eq{RGSU}), the other down-type quarks ($b\rightarrow D$ in \eq{RGSD}) and the other leptons ($\tau\rightarrow E$ in \eq{RGSL}).

The  Yukawa couplings in type I 2HDM are given by, 
\bea
\lambda_f(m_f)= \sqrt{2} \frac{m_f(m_f)}{ v c_\omega}
\eea
where $c_{\omega}=v_1/v$ (with $v=2 M_W/g$). In the SUSY 2HGM  the Yukawa couplings are given by,
\bea
\lambda_U(m_U)&=& \sqrt{2} \frac{m_U(m_U)}{ v  c_\omega s_{\beta_1}} \\
\lambda_D(m_D)&=& \sqrt{2} \frac{m_D(m_D)}{ v c_\omega c_{\beta_1}}\\
\lambda_E(m_E)&=& \sqrt{2} \frac{m_E(m_E)}{ v c_\omega c_{\beta_1}}
\eea 
where $\tan {\beta_1}= v_{u1}/ v_{d1}$.

In the type I 2HDM  all the Yukawa couplings except for the top Yukawa coupling can be ignored. In the SUSY 2HGM in addition to the top Yukawa, the tau and bottom Yukawa couplings also become important  at high values of ${\beta_1}$. We have solved the RGE with the boundary conditions,
\bea
m_t(pole)&=& m_t (m_t) \left(1+\frac{4\alpha_s(m_t)}{3 \pi }\right)= 171.3~{\rm GeV}\\
m_b (M_Z)&=&2.7-3.0~{\rm GeV} ~\cite{Baer}\label{eq:mb}\\
m_\tau (M_Z)&=&1.75~{\rm GeV}
\eea
and $\alpha_s(M_Z)=0.1182$. In \eq{eq:mb} the allowed range corresponds to the Particle Data Group (PDG) range for $m_b (m_b)=4.03-4.37$ GeV~\cite{pdg}. Let $\Lambda_{SC}$ be the strong coupling scale, i.e the energy scale at which either one of the three large Yukawa couplings $\lambda_t, \lambda_b$ or $\lambda_\tau$ becomes larger than $4\pi$. In Fig.~\ref{RGflow} we show how $\Lambda_{SC}$ varies as a function of $c_{\omega}$. We show the  curve for the type I 2HDM case  as well as the curves for the SUSY 2HGM for various values of $\tan {\beta_1}$. In the SUSY 2HGM at high values of $\tan {\beta_1}\gsim 50$  the bottom Yukawa coupling  becomes strong at a lower energy scale than the top  Yukawa coupling.  

We see from the results of this section that if we allow the fermiophobic next generation Higgs doublet too large of a vev, the enhanced Yukawa couplings that are required to make up for the smaller vev of the Higgs boson that the fermions couple to may diverge at a lower scale than desired. For example, if one wishes to preserved supersymmetric gauge coupling unification up to the scale of $\sim 10^{16}\gev$ there are critical values of $\cos\omega$ that cannot be crossed depending on $\tan\beta_1$, which leads to maximum values of the next generation Higgs boson vev.

%%%%%%%%%%%%%%%%%%%%%%%%%%%%%%%%%%%%%%
\section{Signatures at the Large Hadron Collider}
\label{secColl}
In this section we identify processes that can provide signatures of next generation Higgs bosons at the LHC. While our numerical results have been obtained for the type I 2HDM our basic conclusions are true for next generation Higgs bosons in general.
\subsection{Dominant decay modes}

\begin{table}[t]
\centering
\begin{tabular}{c c c }
\hline \hline
Set & Input Parameters &  $\Delta\rho$\\
\hline
A   &$m_H=120$ GeV, $m_h=50$ GeV, $m_A=150$ GeV,  & $0.0010$ \\
    &$m_{H^+}=200$ GeV,$s_\omega=0.1$, $s_\alpha=0.2$ ($\sin \delta=0.10 $)& \\
    &  & \\                                                              
B   &$m_H=120$ GeV, $m_h=70$ GeV, $m_A=180$ GeV,  & $0.0003 $  \\
    &$m_{H^+}=200$ GeV,$s_\omega=0.1$, $s_\alpha=0.2$ ($\sin \delta=0.10 $)&\\
    &&\\
C   &$m_H=120$ GeV, $m_h=100$ GeV, $m_A=200$ GeV,  & $<0.0001 $   \\
    &$m_{H^+}=200$ GeV,$s_\omega=0.1$, $s_\alpha=0.3$ ($\sin \delta=0.20 $)&\\
\hline
\end{tabular}
\caption{Example parameter sets. The values of $\Delta\rho$ are computed in these 2HDMs with respect to the SM value with Higgs mass of $120\gev$.}
\label{tab:input}
\end{table}

We will now compute the branching ratios of the various decay modes of the neutral Higgs bosons in the type I 2HDM. The relevant Feynman rules can be found in Ref.~\cite{ferm1}.

Fig.~\ref{BR}(a) shows the branching ratios for decay of the pseudoscalar $A$ (see \eq{eq:type I potential}). The decay modes $ A\rightarrow hh$, $ A\rightarrow HH$, $ A\rightarrow WW$ and $ A\rightarrow ZZ$ are not allowed by symmetry as $A$ is a pseudoscalar. As shown in the figure, when allowed kinematically, the branching ratio for $ A\rightarrow Zh$ is nearly unity.

The SM-like Higgs $H$  decays mainly via the modes $H\rightarrow hh/AA$ even if modes like $ H\rightarrow WW$ and $ H\rightarrow ZZ$ are kinematically allowed. Fig.~\ref{BR}(b) shows the branching ratios of $H$ in potential A  with a light $h$.

The branching ratios of $h$ are very similar to that of a SM Higgs for small values of $m_h$ when decay modes like $ h\rightarrow ZA$ and $ h\rightarrow AA$ are not kinematically allowed. For $m_h >2 m_A$, the decay mode $ h\rightarrow AA$ overwhelms all other modes including $ h\rightarrow ZA$, $ h\rightarrow WW$ and $ h\rightarrow ZZ$. A very interesting limit is $\alpha \rightarrow 0$, $\omega \rightarrow 0$. In this limit $h$ becomes both fermiophobic and bosophobic (the tree level coupling of $h$ to fermions is proportional to $-s_{\alpha}/c_{\omega}$ and the coupling to vector bosons is proportional to $\sin \delta$) and the dominant decay mode becomes $h \rightarrow \gamma\gamma$. This case has been dealt with in detail in Ref.~\cite{ferm1} and~\cite{ferm2}. 

\subsection{The $pp\rightarrow Z h(b\bar{b})h(b\bar {b}) $ signal and choice of input parameters}

As we noted in section~\ref{smallv2} indirect constraints put an upper bound on $v_2$  in the type I 2HDM and this not only implies a small $\omega$ but also suggests a small $\alpha$ and hence suggests a small $\delta=|\alpha-\omega|$. A small vev for the fermiophobic Higgs doublets is also required for high scale perturbativity as we saw in the last section. If $\delta$ is small  a light $h$ can satisfy the constraints due to the $e^+e^-\rightarrow hZ$ at LEP (Fig.~\ref{sin2delta}). The pseudoscalar $A$ must then be chosen sufficiently heavy to satisfy the LEP constraints from $e^+e^-\rightarrow hA$ (see section~\ref{collider}).

\begin{figure}[t]
%\hspace{-0.5 in}
\begin{center}
\includegraphics[width=1\columnwidth]{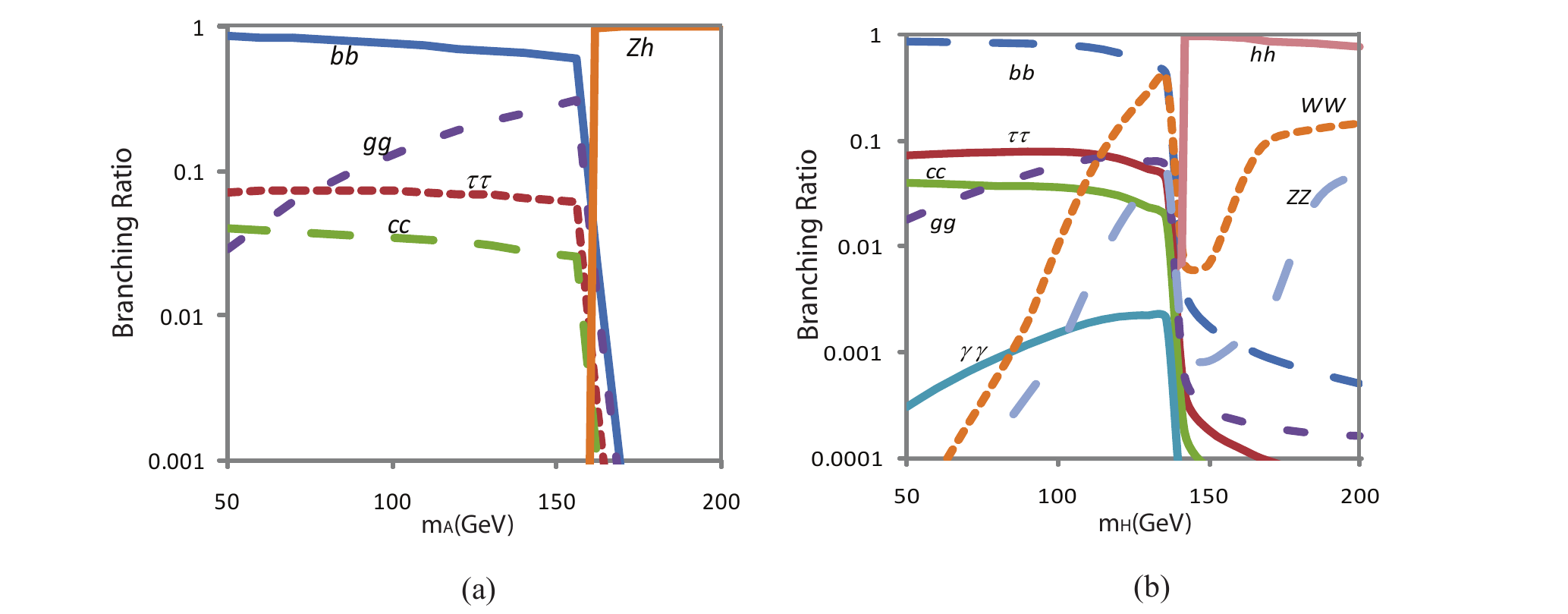}
\end{center}
\caption{(a) Branching ratios of the various decay modes   of the pseudoscalar $A$ with $m_H=120$ GeV, $m_h=70$ GeV,
$m_{H^+}=150$ GeV, $s_\alpha=0.2$ and $s_\omega=0.1$ in any 2HDM and (b) branching ratios of the SM-like Higgs $H$ in potential A with $m_A=180$ GeV, $m_h=70$ GeV, $m_{H^+}=150$ GeV, $s_\alpha=0.2$ and $s_\omega=0.1$.}
\label{BR}
\end{figure}

Another implication of a small $\delta$ is that unlike the process $pp\rightarrow hZ$, which is highly suppressed (by $\sin^2 \delta$ in 2HDMs), the processes $pp\rightarrow Ah$ and $pp\rightarrow ZH$ are only mildly suppressed (by $\cos^2 \delta$ in 2HDMs). In the first case, if allowed kinematically, $A$  predominantly decays as $A\rightarrow Zh$ (Fig.~\ref{BR}(a)) and in the second case $H$  predominantly decays as $H\rightarrow hh/AA$ (Fig.~\ref{BR}(b)) if allowed to kinematically so that both processes can lead to the same final state $Z h(b\bar{b})h(b\bar {b})$.\footnote{Note, however, that in a 2HDM with potential B, in the limit $\alpha,\omega\rightarrow 0$ we get $m_h^2\rightarrow m_A^2$ (see \eq{pseudoB} and \eq{hmassB}) so that $A\rightarrow Zh$ would not be kinematically allowed in this limit. This issue does not arise for potential A or the general potential in \eq{genpot}.}   

\begin{figure}[t]
%\hspace{1 cm}
\begin{center}
\includegraphics[width=0.7\columnwidth]{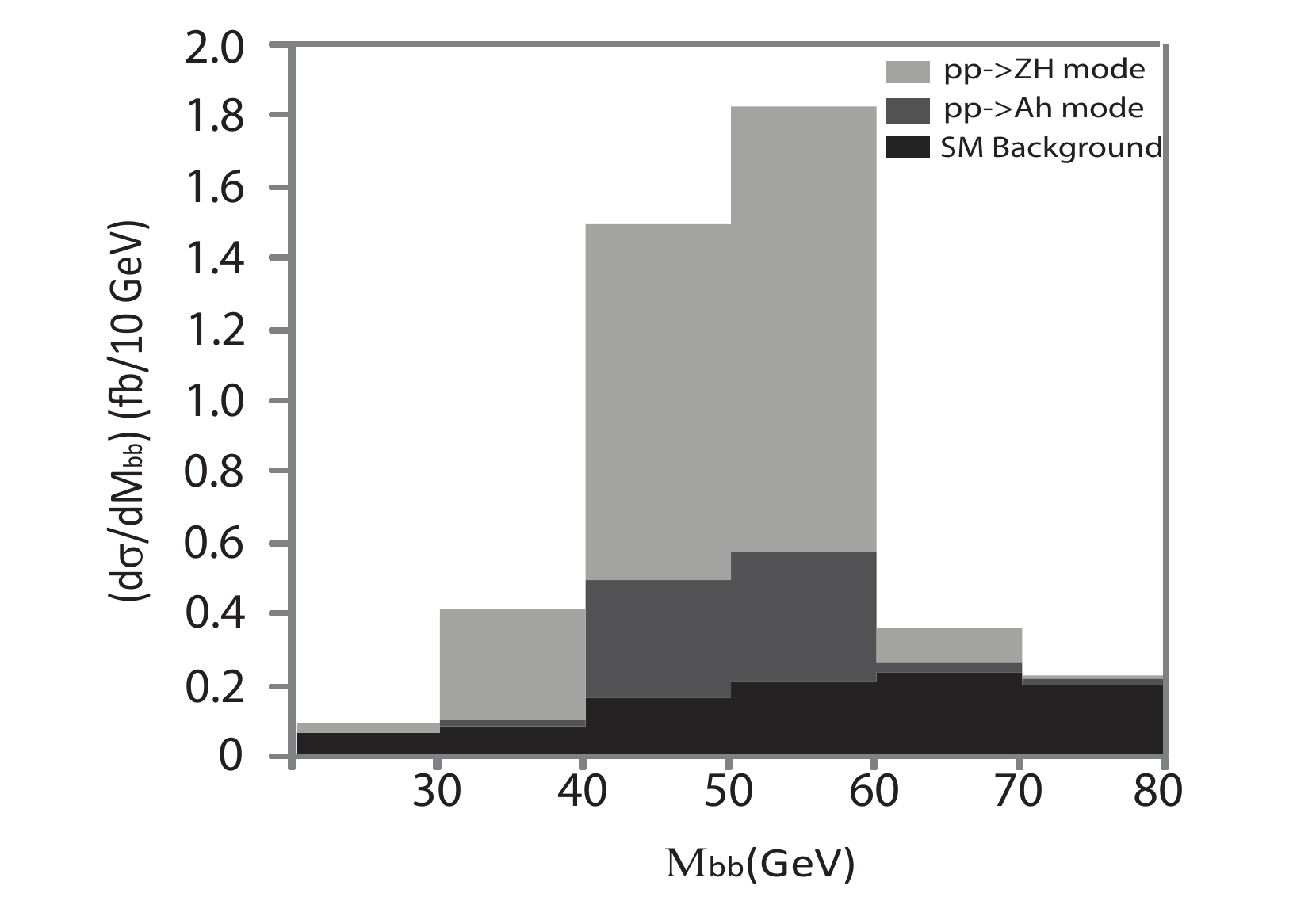}
\end{center}
\caption{The 2$b$ invariant mass spectrum $M_{bb}$ for the signal on top of the SM background for the input parameters in  Set A (see Table~\ref{tab:input}), showing a peak at $m_h=50$ GeV. Two processes contribute to the signal (1) $pp\rightarrow A(Zh)h \rightarrow Z(l^+l^-)Z b\bar{b}b\bar {b}$ and  (2) $pp \rightarrow ZH(hh) \rightarrow Z(l^+l^- )Z b\bar{b}b\bar {b}$ . The cuts applied are those in eqs.~(\ref{cuts1})-(\ref{cuts2}) and  the center of mass energy has been taken to be 14 TeV. The branching ratio $B(H\rightarrow hh)\approx 1$. The reconstruction efficiency of the lepton pair and the jets and the $b$-tagging efficiency have all been taken to be unity at this stage.}
\label{setA2b}
\end{figure}

\begin{figure}[t]
\begin{center}
\includegraphics[width=0.7\columnwidth]{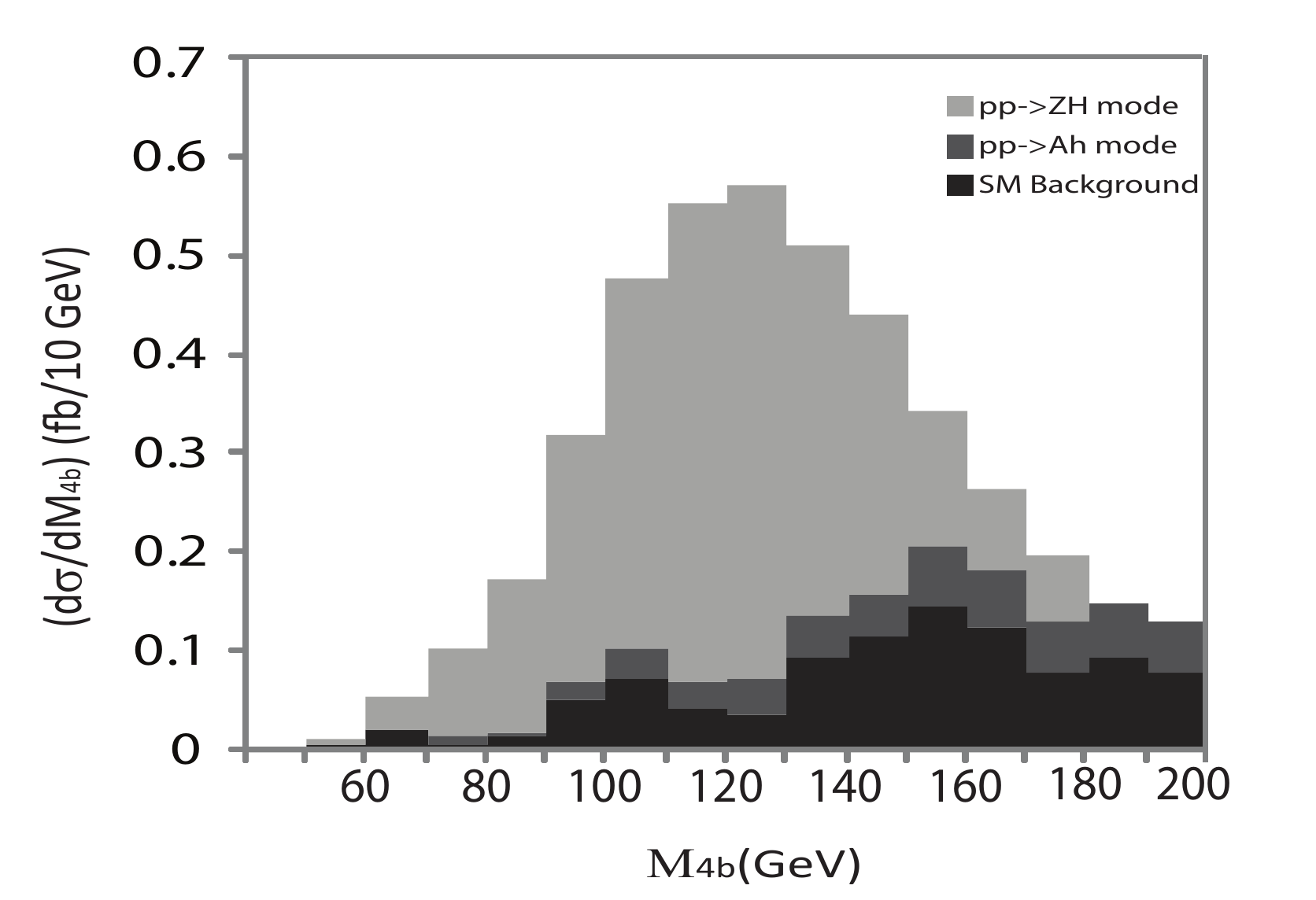}
\end{center}
\caption{The 4$b$ invariant mass spectrum $M_{4b}$ for the signal on top of the SM background for the input parameters in  Set A (see Table~\ref{tab:input}). Two processes contribute to the signal (1) $pp \rightarrow  A(Zh)h\rightarrow Z(l^+l^- )Z b\bar{b}b\bar {b}$ and  (2) $pp \rightarrow  ZH(hh) \rightarrow Z(l^+l^- )Z b\bar{b}b\bar {b}$ .  The  $pp \rightarrow  ZH(hh)$ mode gives rise to a peak at $m_H=120$ GeV. We have smeared the 4$b$ invariant mass assuming an experimental resolution of 20$\%$ of $m_H$  (24 GeV in this case) for the reconstructed peak. The cuts applied are those in eqs.~(\ref{cuts1})-(\ref{cuts2}) and the center of mass energy has been taken to be 14 TeV. The branching ratio $B(H\rightarrow hh)\approx 1$. The reconstruction efficiency of the lepton pair and the jets and the $b$-tagging efficiency have all been taken to be unity at this stage.}
\label{setA4b}
\end{figure}

\begin{figure}[t]
%\hspace{-1 in}
\begin{center}
\includegraphics[width=0.75\columnwidth]{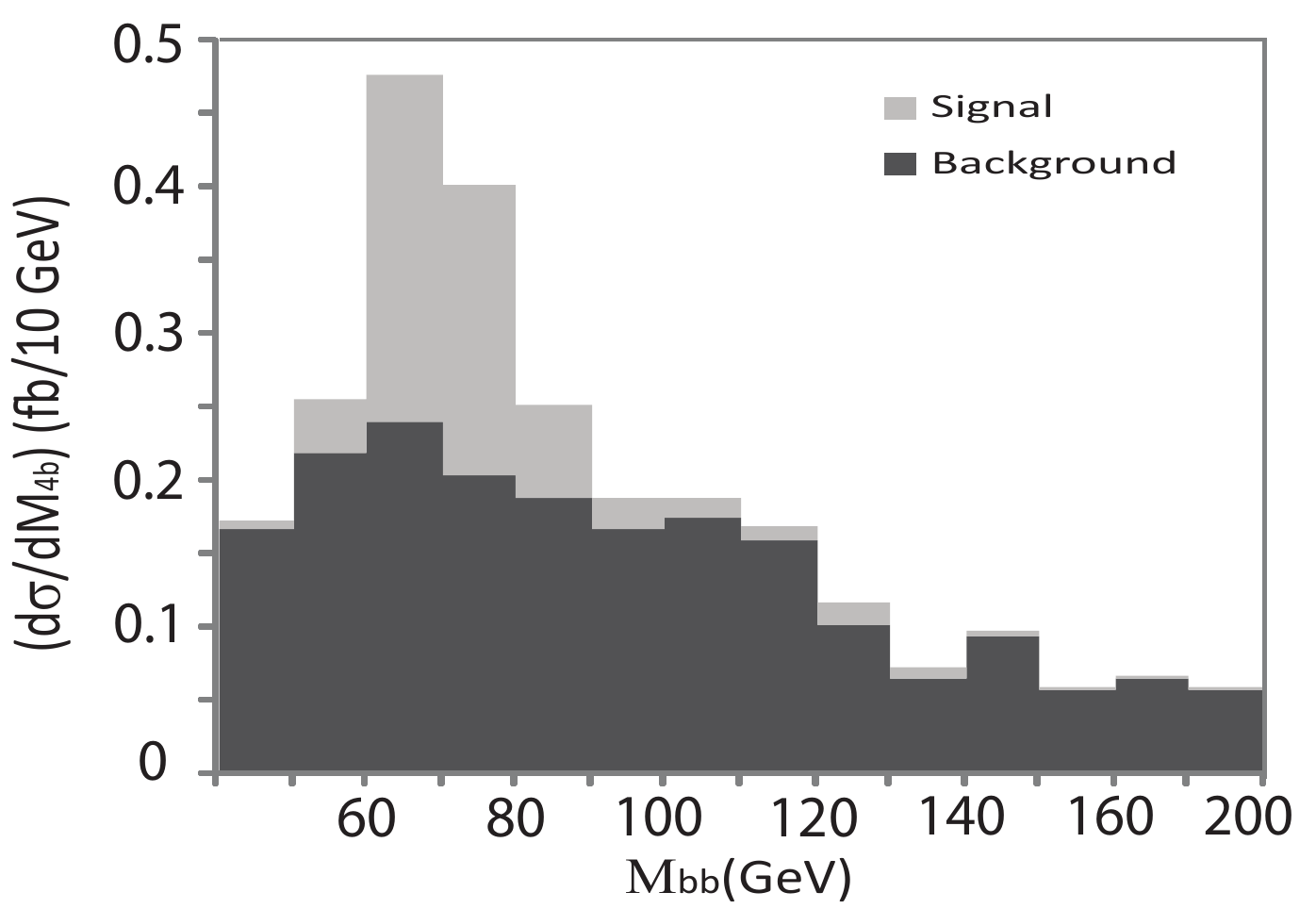}
\end{center}
\caption{Invariant mass spectrum $M_{bb}$ for the signal on top of the SM background for the input parameters in  Set B (see Table~\ref{tab:input}), showing a peak at $m_h=70$ GeV. The signal process is $pp \rightarrow  A(Zh)h \rightarrow Z(l^+l^- )Z b\bar{b}b\bar {b}$ . The cuts applied are those in eqs.~(\ref{cuts1})-(\ref{cuts2}) and  the center of mass energy has been taken to be 14 TeV. The reconstruction efficiency of the lepton pair and the jets and the $b$-tagging efficiency have all been taken to be unity at this stage. }
\label{setB}
\end{figure}

\begin{figure}[t]
%\hspace{0.8 in}
\begin{center}
\includegraphics[width=0.75\columnwidth]{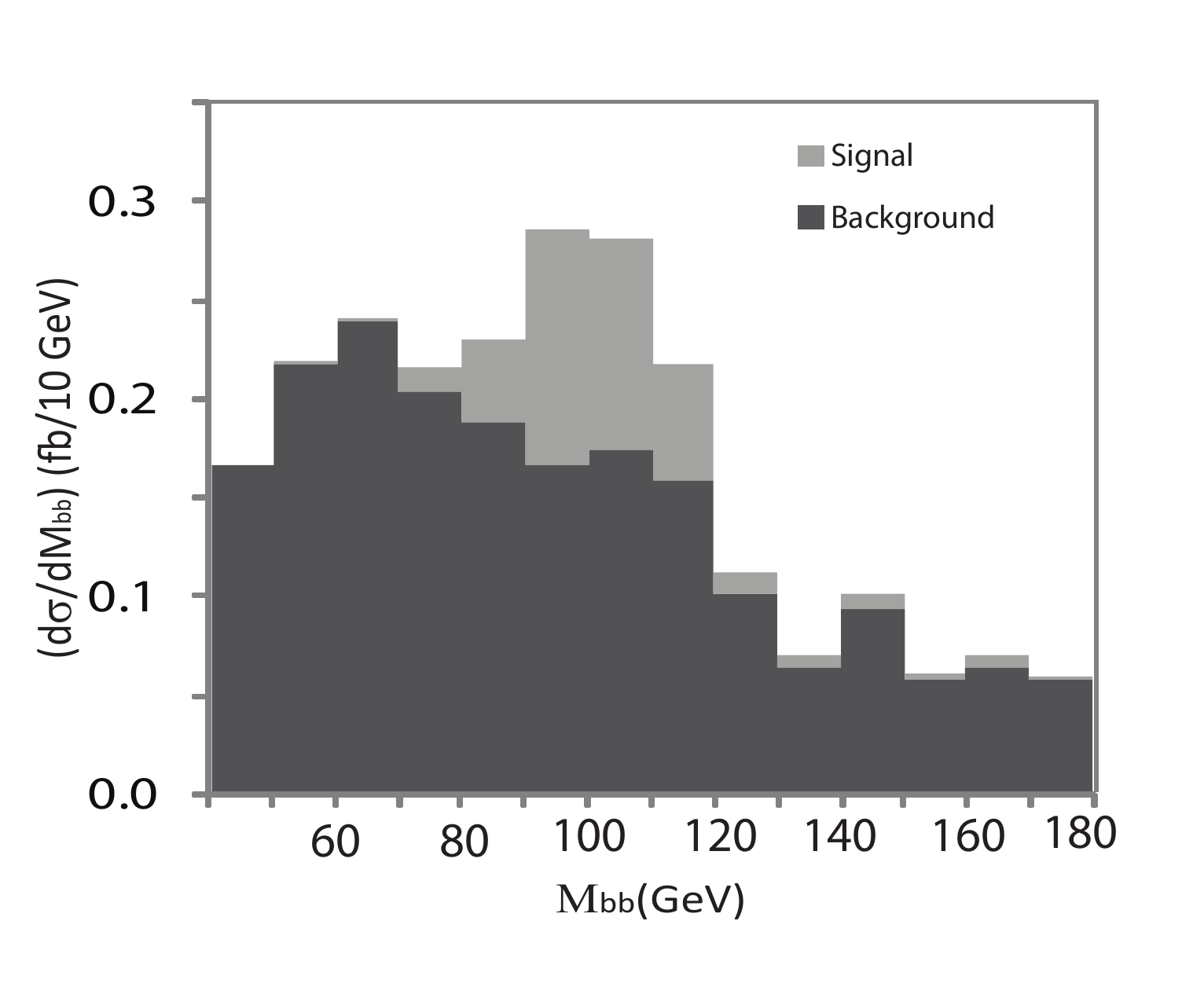}
\end{center}
\caption{Invariant mass spectrum $M_{bb}$ for the signal on top of the SM background for the input parameters in  Set C (see Table~\ref{tab:input}), showing a peak at $m_h=100$ GeV. The signal process is  $pp \rightarrow  A(Zh)h \rightarrow Z(l^+l^- )Z b\bar{b}b\bar {b}$. The cuts applied are those in eqs.~(\ref{cuts1})-(\ref{cuts2}) and  the center of mass energy has been taken to be 14 TeV. The reconstruction efficiency of the lepton pair and the jets and the $b$-tagging efficiency have all been taken to be unity at this stage.
}
\label{setC}
\end{figure}

\begin{table}[t]
\centering
\begin{tabular}{c c c c }
\hline 
\hline
    & Cross Section         &  Cross Section         & Detector level Cross-\\
Set & with the cuts   &  with the cuts   &Section ($\sigma_{eff}$ in \eq{eff})\\
                     &    in eqs.~(\ref{cuts1})-(\ref{cuts1c}) (fb) &  in eqs.~(\ref{cuts1})-(\ref{cuts2}) (fb)   &with the cuts \\
  &&&in eqs.~(\ref{cuts1})-(\ref{cuts2}) \\
\hline
                     &     &     & \\
\textbf{Set A }      &     &     & \\
$pp\rightarrow Ah$   &1.2 & 0.8& 0.1\\
$pp\rightarrow ZH$   &3.6 & 2.7&0.4 \\
   \textit{Total}    &4.8 & 3.5& 0.5\\
                     &     &     & \\ 
\textbf{Set B}       &     &     & \\ 
$pp\rightarrow Ah$   &0.9 & 0.6&0.09 \\
                     &     &     & \\
 \textbf{Set C}      &     &     & \\
$pp\rightarrow Ah$   &0.6 & 0.4& 0.06\\
                     &     &     & \\
SM Background        &9.9 & 2.6&0.4 \\ 
                     &     &     &\\
\hline
\end{tabular}
\caption{Signal and background cross sections for $pp\rightarrow Zh(b\bar {b})h(b\bar{b}$) at LHC. The center of mass energy has been taken to be 14 TeV and the acceptance cuts are those mentioned in the top row. }
\label{tab:results}
\end{table}  

Table~\ref{tab:input} shows three different sets of input parameters for which we perform simulations to compute the $pp\rightarrow Z h(b\bar{b})h(b\bar {b}) $ signal. Set A receives contributions from both the  $pp\rightarrow Ah$ and $pp\rightarrow ZH$ processes whereas all the other sets receive contributions from the $pp\rightarrow Ah$ process only. We have also computed $\Delta \rho$ for all the data sets in Table~\ref{tab:input} to show that this constraint is met. As for the other constraints, we have marked these parameter sets in Fig.~\ref{bsgamma} and Fig.~\ref{sin2delta} to show  that our parameter sets satisfy the $b\rightarrow s \gamma$ constraint and the $e^+e^-\rightarrow hZ$ constraint respectively.

As mentioned in section~\ref{collider} a light $A$ is experimentally less constrained than a light $h$. Although we will not perform simulations for the process $pp\rightarrow A(b \bar{b})h$ followed by $h\rightarrow ZA(b \bar{b})$ the analysis of the process would be very similar to the process we will consider. This process would be important if $h\rightarrow ZA$ is kinematically allowed but $h\rightarrow AA$ is not kinematically allowed. If $m_H>2 m_A$ the process $pp \rightarrow ZH(AA)\rightarrow Z b\bar{b}b\bar {b}$ will be a very important signature. Although we will not perform a  simulation for such a scenario, we will compute the contribution to the cross section of the process  $pp \rightarrow ZH(hh)\rightarrow Z b\bar{b}b\bar {b}$ for Set A (this will be a part of the net cross section). This is again expected be very similar to the case when $A$, rather  than $h$, is the lighter scalar that $H$ decays into. 

It should be clear that all the arguments that have led us to the $pp \rightarrow A(Zh)h\rightarrow Z h(b\bar{b})h(b\bar {b})$ signal are valid not only for the type I 2HDM but for any multi-Higgs theory having scalar doublets that do not couple to fermions.  It is easiest to satisfy the constraints from charged Higgs loop contributions (as in the $b\rightarrow s \gamma$ process) and high scale perturbativity  if the vevs of these fermiophobic doublets are small. Let $h$ and $A$ be mass eigenstates that contain mostly the CP-even and CP-odd neutral components of such a doublet respectively. A small vev of this doublet would imply that the $ZZh$ coupling strength is  small, but the $ZAh$  coupling strength would not get suppressed.  

%%%%%%%%%%%%%%%%%%%%%%%%%%%%%%%%%
\subsection{Signal and Background cross section at LHC}

We now present the results of the simulations we performed for the $pp \rightarrow Z h(b\bar{b})h(b\bar {b})$ signal. The analyses we present are new; however, there are related studies that we will point out to the reader in our Set A analysis. All our results are independent of the choice of 2HDM potential except for the $H\rightarrow hh$ contribution to the signal in parameter set A that depends on $B(H\rightarrow hh)$ which is model dependent. $B(H\rightarrow hh)$  has been taken to be equal to its value in potential A which is nearly  1. We used MADGRAPH~\cite{Alwall} to generate signal and background events at 14 TeV center of mass energy for the process  $pp \rightarrow Z h(b\bar{b})h(b\bar {b})$ for the different parameter sets in Table~\ref{tab:input}. We then decayed the $Z$ to $l^+ l^- ~(l=e,\mu)$  using the DECAY software in the MADGRAPH package. Note that we have ignored any contribution to the SM background from final states with lepton pairs not produced in $Z$ decay that have invariant mass close to $M_Z$ nevertheless.  The following basic selection cuts have been applied using MadAnalysis, 
\bea
p_{T}(b,l)  &>& 15~{\rm GeV}
\label{cuts1}
\\
|\eta_{b,l}| &<& 2.5
\\
\Delta R_{bp}  &>& 0.4.
\label{cuts1c}
\eea
where $\Delta R=\sqrt{(\Delta \eta)^2+(\Delta\phi)^2}$ and $p$ is any parton (i.e. quark or lepton) in the process.
The background cross section with these cuts is 9.9 fb. To reduce the background further we use the fact that two $b\bar{b}$ pairs  have the same invariant mass in the signal up to experimental resolution. Detector simulations of scalars decaying into $b$-pairs~\cite{cav}  find  that because of  detector effects like calorimeter energy resolution, electronic noise,  and physics effects like final state radiation, energy loss outside cone and semi leptonic decays, only about 85$\%$ of the events register  di-\textit{b}-jet invariant masses within 20$\%$ of the true value. To simulate this effect we  smear the invariant masses  according to a Gaussian distribution such that 85$\%$ of the events lie within 20$\%$ of the mean. There are three ways to divide the four $b$-quarks (say \textit{abcd}) into two pairs (\textit{ab cd}, \textit{ac bd} and \textit{ad bc}). The combination that gives the invariant masses of the two pairs (after smearing) closest to each other has been considered (note that in reality the experimental uncertainties in the invariant masses of the six possible $b$-pairs are not uncorrelated as assumed here). Let $M_{bb1}$ and $M_{bb2}$ be the smeared invariant masses of the two $b$ pairs thus selected, and let $M_{bb}$ be the mean of these two numbers. We impose the following cut in addition to those in \eq{cuts1},
\bea
|M_{bb1}-M_{bb2}|<0.2~M_{bb}.
\label{cuts2}
\eea
With this additional cut the background is reduced from 9.9 fb to 2.6 fb whereas the signal is only reduced to  about 70$\%$ of the value with only the cuts in eqs.~(\ref{cuts1})-(\ref{cuts1c}). We provide the cross sections for the different parameter sets and the background in Table~\ref{tab:results}. Note that the background cross section can be further reduced by requiring $M_{bb}$ to be in a certain mass window around $m_H$

In Fig.~\ref{setA2b} we plot the 2$b$ invariant mass spectrum for Set A obtained for the events passing  the cuts in eqs.~(\ref{cuts1})-(\ref{cuts2}). For Set A the contribution to the cross section of $pp \rightarrow Z h(b\bar{b})h(b\bar {b})$ comes from two different processes $pp \rightarrow A(Zh)h\rightarrow Z b\bar{b}b\bar {b}$ and $pp \rightarrow ZH(hh)\rightarrow Z b\bar{b}b\bar {b}$. The branching ratio $B(H\rightarrow hh)$ has been taken to be equal to its potential A value which is nearly 1. As shown in Fig.~\ref{setA2b} the contribution from  $pp \rightarrow ZH(hh)$ mode is quite large.  This contribution  also gives rise to a peak in the 4\textit{b} invariant mass spectrum as shown in Fig.~\ref{setA4b}. We have smeared the 4$b$ invariant mass assuming an experimental resolution of 20$\%$ of $m_H$  (24 GeV in this case) for the reconstructed peak. We expect very similar results if $A$ rather than $h$ is the scalar $H$ decays to. The signal from $pp \rightarrow ZH(AA)\rightarrow Z b\bar{b}b\bar {b}$ will be a very promising signature for the type I 2HDM, especially because the constraints on a light $A$ are rather weak as discussed in section~\ref{collider}. Similar analyses for $H\rightarrow AA$ have been done before (in Ref.~\cite{ch} the final state $lA(b\bar{b})A(b\bar {b})$ has been considered while in Ref.~\cite{han}  the final state $W(E^{miss}_T l)A(b\bar{b})A(b\bar {b})$ has been considered). These papers, however, do not apply the cut in \eq{cuts2} which leads to improved significance of the signal.  

Fig.~\ref{setB},~\ref{setC} shows the 2$b$ invariant mass spectrum for the other parameter sets. Only the process $pp \rightarrow A(Zh)h\rightarrow Z b\bar{b}b\bar{b}$ contributes to the cross section in these cases.

To get the  cross section we expect the detectors to effectively measure we must multiply by the  efficiency of reconstruction of a lepton pair  and that of four jets. These efficiencies depend on kinematical quantities like $p_T$ and $\eta$. We take an average value 0.8 for lepton pair reconstruction efficiency (see pgs.\ 72-92, pgs.\ 210-223 in Ref.~\cite{Atl}) and 0.9 for  reconstruction efficiency of a jet (see pgs.\ 286-287 in Ref.~\cite{Atl}). We also require  that at least three of the jets are $b$-tagged which gives an overall $b$-tagging efficiency equal to $ {4\choose 3} \epsilon_b^3 (1-\epsilon_b)+\epsilon_b^4$ where the $b$-tagging efficiency for  single jet  $\epsilon_b\approx 0.5$~\cite{Atl}.  Putting it all together we get,
\bea
\sigma^{eff}=0.16~\sigma.
\label{eff}
\eea
This equation is applicable for both the cross section and the background. The effective signal and background cross sections also appear in Table~\ref{tab:results}.

We have not computed the contribution due to mistagging of $c$-jets or other light jets. This will decrease the signal significance somewhat as mistagging is expected to have an appreciable contribution only to the SM background and not the signal. This is because the signal cross section is proportional to the square of the branching ratio of $h$ to the mistagged quarks but the branching ratio of $h$ to quarks other than $b$-quarks is much smaller. In Ref.~\cite{han} the signal and background cross sections for the process $pp\rightarrow WH \rightarrow W(E^{miss}_T l)A(b\bar{b})A(b\bar {b})$ have been computed including mistagging effects. A rough estimate of the cross sections they obtain can be made from Fig.~5 in their paper. Such an estimate shows that the background cross section they obtain due to mistagged quarks  is about a third of the contribution due to correctly tagged $b$ quarks.  This mistag background is highly dependent upon the details of detector performance issues that will be sorted out in the course of the LHC runs. We do not expect the addition of these backgrounds to substantively change the discovery capability that we have presented, especially since they are unlikely to peak at the Higgs mass $m_h$ in the di-jet invariant mass spectrum.

\section{Conclusion and Discussion}

In conclusion, we have argued that next generation Higgs bosons should be viewed as generic possibilities in string theory model building, and illustrated this viewpoint with recent developments in string phenomenology.  We presented a generalized theorem for the structure of Higgs couplings to SM fermions that automatically avoids problematic tree-level flavor changing neutral currents that are induced by new Higgs boson exchanges. Our viewpoint is that the interaction rules of this theorem are too restrictive to be satisfied without a principle. In the case of the two-Higgs doublet model of supersymmetry, the principle is holomorphy. Additional Higgs bosons added in any other context, such as more Higgs doublet pairs in supersymmetry or simply another Higgs boson in the SM, requires a strong discrete symmetry or selection rule. This can be contemplated within effective field theories, for example by $\Phi\to -\Phi$ $Z_2$ symmetry, or within string theory by algebraic selection rules that may not totally forbid the unwanted couplings but can approximate zero, as was the case in the work of~\cite{Ambroso:2008kb}.

Next we considered various constraints that these theories must face. For example, although tree-level flavor changing neutral currents induced by neutral Higgs exchanges may be satisfied, loop-level ones induced by the charged Higgs boson may not be. The $b\to s\gamma$ transition is quite constraining to exotic Higgses that get large vacuum expectation values, since they steal vev from the Higgs that couples to fermions, thereby raising those fermion Yukawa coupling magnitudes to dangerous levels. Even normal Cabibbo-Kobayashi-Maskawa (CKM) mixings can create too-large amplitude shift in that case compared to what experiment allows, and therefore the parameter space is not completely open and limits are derived on the exotic Higgs vev  as a function of the charged Higgs boson mass. 

We computed the spectrum of Higgs boson states with a next generation in the supersymmetric and non-supersymmetric context. Within supersymmetry we showed that there is the prospect of slightly raising the tree-level CP even Higgs boson mass with respect to the MSSM in the smaller $\tan\beta$ region.  The effects are largest when the next generation Higgs boson has a large vacuum expectation value. The large Yukawa couplings that  are present when an exotic, fermiophobic Higgs doublet takes a large vev can alter the domain of perturbativity of the theory. We showed both in the SM context and the supersymmetry context that the top quark Yukawa coupling could develop a Landau pole well below the Planck scale. A low-scale Landau pole would preclude the existence of a perturbative theory description of unification at the high scale, and these results must be taken into account when considering a next generation of Higgs bosons.

Finally, we investigated the phenomenology of the exotic Higgs sectors at the LHC. Multi-Higgs boson phenomenology within supersymmetry is a mature topic; however, the fermiophobic next generation Higgs boson layer to the theory has not been considered in depth. The salient new features are similar to SM phenomenology  with an additional fermiophobic Higgs doublet. Thus, we  discussed collider physics possibilities within that less complex framework.  

A particularly interesting possibility is the production of $hA$ at the LHC, followed by $A\to hZ$. Assuming $h$ is rather light, say less than $\sim 150\gev$, we can expect the largest branching fraction of $h$ decays to be to $b\bar b$.  For good distinction from background we can also demand the $Z$ decay to leptons $l^+l^-$. Thus, the signal becomes $4b+2l$. Background becomes particularly limited when we require at least three $b$-quark tags and that the four jets reconstruct two equal mass resonances.  We propose that as a search strategy for this case, and show that there are good prospects for the LHC to find this signal. Discovery would be an indication of next generation Higgs bosons.

\bigskip\bigskip
\noindent
{\bf Acknowledgments:} We thank A. Ali, B. Campbell, H. Haber, G. Kane, J. Kumar, S. Martin, M. Misiak, B. Ovrut, A. Pierce, J. Qian and F. Quevedo for helpful conversations. This work is supported in part  by CERN, the U.S. Department of Energy and  the European Commission under the contract  ERC advanced grant 226371 `MassTeV'.

\end{document}